\documentclass[iop]{emulateapj}
\usepackage{natbib}
\usepackage{graphicx,color,rotating}
\usepackage{footnote,lineno}
\usepackage{ulem} 
\usepackage{xspace}
\usepackage{graphicx,amssymb,amsmath,amsfonts,times,hyperref}

\def \aap  {A\&A}

\def \apj  {ApJ}
\def \apjs  {ApJS}
\def \apjl  {ApJL}

\newcommand{\fermipy}{\texttt{Fermipy}\xspace}

\shorttitle{Characterizing the population of pulsars in the
  inner Galaxy with the  {\it Fermi} Large Area Telescope.}

\shortauthors{The Fermi-LAT Collaboration}

\begin{document}
\title{Characterizing the population of pulsars in the
  inner Galaxy with the  {\it Fermi} Large Area Telescope.}

\author{
M.~Ajello\altaffilmark{1}, 
L.~Baldini\altaffilmark{2}, 
J.~Ballet\altaffilmark{3}, 
G.~Barbiellini\altaffilmark{4,5}, 
D.~Bastieri\altaffilmark{6,7}, 
R.~Bellazzini\altaffilmark{8}, 
E.~Bissaldi\altaffilmark{9,10}, 
R.~D.~Blandford\altaffilmark{11}, 
E.~D.~Bloom\altaffilmark{11}, 
E.~Bottacini\altaffilmark{11}
% ...
}
\altaffiltext{1}{Department of Physics and Astronomy, Clemson University, Kinard Lab of Physics, Clemson, SC 29634-0978, USA}
\altaffiltext{2}{Universit\`a di Pisa and Istituto Nazionale di Fisica Nucleare, Sezione di Pisa I-56127 Pisa, Italy}
\altaffiltext{3}{Laboratoire AIM, CEA-IRFU/CNRS/Universit\'e Paris Diderot, Service d'Astrophysique, CEA Saclay, F-91191 Gif sur Yvette, France}
\altaffiltext{4}{Istituto Nazionale di Fisica Nucleare, Sezione di Trieste, I-34127 Trieste, Italy}
\altaffiltext{5}{Dipartimento di Fisica, Universit\`a di Trieste, I-34127 Trieste, Italy}
\altaffiltext{6}{Istituto Nazionale di Fisica Nucleare, Sezione di Padova, I-35131 Padova, Italy}
\altaffiltext{7}{Dipartimento di Fisica e Astronomia ``G. Galilei'', Universit\`a di Padova, I-35131 Padova, Italy}
\altaffiltext{8}{Istituto Nazionale di Fisica Nucleare, Sezione di Pisa, I-56127 Pisa, Italy}
\altaffiltext{9}{Dipartimento di Fisica ``M. Merlin" dell'Universit\`a e del Politecnico di Bari, I-70126 Bari, Italy}
\altaffiltext{10}{Istituto Nazionale di Fisica Nucleare, Sezione di Bari, I-70126 Bari, Italy}
\altaffiltext{11}{W. W. Hansen Experimental Physics Laboratory, Kavli Institute for Particle Astrophysics and Cosmology, Department of Physics and SLAC National Accelerator Laboratory, Stanford University, Stanford, CA 94305, USA}
% ...


\begin{abstract}
  An excess of $\gamma$-ray emission from the Galactic Center (GC)
  region with respect to predictions based on a variety of
  interstellar emission models and $\gamma$-ray source catalogs has
  been found by many groups using data from the {\it Fermi} Large Area
  Telescope (LAT). Several interpretations of this excess have been
  invoked. In this paper we search for members of an unresolved
  population of $\gamma$-ray pulsars located in the inner Galaxy that 
  are predicted by the interpretation of the GC excess as being due to a
  population of such sources.
  We use cataloged LAT sources to derive criteria that efficiently
  select pulsars with very small contamination from blazars.  We
  search for point sources in the inner $40^\circ\times40^\circ$
  region of the Galaxy, derive a list of approximately 400 sources,
  and apply pulsar selection criteria to extract pulsar candidates
  among our source list. We performed the entire data analysis
  chain with two different interstellar emission models (IEMs), and found
  a total of 135 pulsar candidates, of which 66 were selected with 
  both IEMs.
\end{abstract}

\maketitle

\section{Introduction}
The Large Area Telescope (LAT) on board the {\it Fermi} Gamma-ray Space Telescope
has been operating since 2008.  It has produced the most
detailed and precise maps of the $\gamma$-ray sky and collected
more than 200 million extraterrestrial $\gamma$ rays in the energy
range 0.05--2000~GeV.

The region toward the Galactic Center (GC) is the brightest direction
in LAT maps. Along this line of sight (l.o.s.) $\gamma$
  rays originate primarily in diffuse processes: interactions of
primary cosmic-ray (CR) nuclei with the interstellar gas,
bremsstrahlung scattering of CR electrons and positrons with
interstellar gas, and inverse Compton scattering of photons from
interstellar radiation fields.  The LAT also detects individual
sources such as pulsars, compact binary systems, supernova remnants,
and blazars.  In the last seven years many groups analyzing LAT data
have reported the detection of an excess of $\gamma$-ray emission at
GeV energies with an extent of about $20^\circ$ from the GC (we will
refer to this as the GC excess).

The GC excess is found with respect to predictions based
  on a variety of interstellar emission models (IEMs), point source
catalogs, and selections of LAT
data~\citep[e.g.,][]{2009arXiv0910.2998G,2012PhRvD..86h3511A,
  2013PDU.....2..118H,2013PhRvD..88h3521G,Abazajian:2014fta,
  Calore:2014xka,2016PDU....12....1D}.
This excess is well modeled with a spherically symmetric generalized
Navarro-Frenk-White
(NFW)~\citep{1996ApJ...462..563N,1998ApJ...502...48K} density profile
with index $\alpha =1.25$, $\rho(r) = \rho_0/( r/r_s (1+r/r_s))^\alpha$,
 and its spectral energy distribution (SED) in the
inner $10^{\circ}$ from the GC is peaked at a few GeV with an
intensity that is approximately one tenth of the total $\gamma$-ray
intensity.

The {\it Fermi}-LAT Collaboration has performed an analysis
using 5.2 years of the Pass 7 reprocessed data in the energy range 1 to 100~GeV for the
$15^{\circ}\times15^{\circ}$ region around the GC.
This analysis constructed four dedicated IEMs 
and produced a point-source catalog (designated 1FIG) which
includes 48~sources detected in each of the four IEMs with a Test Statistic (TS) larger than
25 \footnote[1]{The TS is defined as twice the difference in maximum log-likelihood 
between the null hypothesis (i.e., no source present)
and the test hypothesis: $TS = 2 ( \log\mathcal{L}_{\rm test} -
\log\mathcal{L}_{\rm null} )$~\citep[see, e.g.,][]{1996ApJ...461..396M}.}
~\citep{2016ApJ...819...44A}.  

Recently, the {\it Fermi}-LAT Collaboration published an updated
analysis~\citep{TheFermi-LAT:2017vmf} using data from 6.5 years of observation
and the new Pass 8 event-level analysis~\citep{2013arXiv1303.3514A}.
The Pass 8 event-level analysis significantly improves the acceptance,
direction and energy reconstruction, and enables sub-selection of
events based on the quality of the direction reconstruction.
In this updated analysis further investigations of
  the systematic uncertainties of modeling the diffuse emission region
  were made using a variety of templates for additional diffuse $\gamma$-ray emission components, such as a
data-motivated template for the {\it Fermi}
bubbles~\citep{2010ApJ...724.1044S,2014ApJ...793...64A}, and with an
additional population of electrons used in modeling the central
molecular zone, and with three different point source lists.

These two analyses confirm the existence of the GC excess.  However, the
energy spectrum of the excess is found to depend significantly on the
choice of IEM and source list~\citep{2016ApJ...819...44A,TheFermi-LAT:2017vmf}.

Different interpretations have been proposed to explain the GC excess.  
Its approximately spherical morphology
and energy spectrum are compatible with $\gamma$ rays emitted from a
Galactic halo of dark matter (DM).  This possibility has been studied
in many papers~\citep[e.g.,][]{2009arXiv0910.2998G,Abazajian:2014fta,Calore:2014xka,2016PDU....12....1D} and
the intensity and shape of the GC excess has been found to be
compatible with DM particles with mass 40--60~GeV annihilating
through the $b\bar{b}$ channel with a thermally averaged cross section close to the
canonical prediction for thermal relic DM~\citep[roughly $3\times10^{-26} {\rm cm}^3
{\rm s}^{-1}$, e.g.,][]{Steigman:2012nb}.

However, if DM exists and gives rise to this excess the same particles
should also produce measurable emission from dwarf spheroidal
satellite galaxies of the Milky Way, which are known to be DM
dominated ~\citep{2008Natur.454.1096S}. 
No evidence of such a flux from dwarf galaxies has been
detected so far, and the limits obtained for the annihilation cross
section are in tension with the DM interpretation of the GC
excess~\citep[see][and references therein]{2017ApJ...834..110A}.

Among alternative interpretations proposed are that the GeV excess is
generated by recent outbursts of CR protons interacting with gas via
neutral pion production~\citep{Carlson:2014cwa} or of CR leptons
inverse Compton scattering interstellar
radiation~\citep{Petrovic:2014uda,Cholis:2015dea,Gaggero:2015nsa}.
However, the hadronic scenario predicts a $\gamma$-ray signal that is
significantly extended along the Galactic plane and correlated with
the distribution of gas, which is highly incompatible with the
observed characteristics of the excess \citep{Petrovic:2014uda}.  The
leptonic outburst scenario plausibly leads to a signal that is more
smoothly distributed and spherically symmetric; however, it requires
at least two outbursts to explain the morphology and the intensity of
the excess with the older outbursts injecting more-energetic
electrons. An additional population of supernova remnants near the GC
that steadily injects CRs is also a viable interpretation for the GC
excess~\citep{Gaggero:2015nsa,Carlson:2015ona}.

Recently, evidence of the existence of an unresolved population of
$\gamma$-ray sources in the inner $20^\circ$ of the Galaxy with a
total flux and spatial distribution consistent with the GC excess has
been published by \citet{Lee:2014mza} and \citet{Bartels:2015aea}. These faint
sources have been interpreted as belonging to the Galactic bulge PSR population.  
This interpretation has been investigated, by, e.g.,~\citet{Cholis:2014lta}
who claim that about 60 Galactic bulge pulsars should have been
already present in {\it Fermi}-LAT catalogs, though they may not
yet be firmly identified as PSRs.

Several authors have examined the properties of detected $\gamma$-ray
pulsars (PSRs) and found that the unresolved pulsars in the Galactic
bulge could account for a significant fraction of the GC
excess~\citep{2013MNRAS.436.2461M,Gregoire:2013yta,2014JHEAp...3....1Y,Petrovic:2014xra,2015ApJ...812...15B,OLeary:2015qpx,OLeary:2016cwz}.
Throughout this paper we will use ``PSR'' to refer
specifically to detectable $\gamma$-ray pulsars; i.e., pulsars that
emit $\gamma$-rays and whose $\gamma$-ray beams cross the Earth.
Typically these have spin-down luminosities above the observed
``deathline'' of $\sim 3 \times 10^{33} {\rm \,erg}{\rm
  s}^{-1}$~\citep{2016A&A...587A.109G}. The SEDs of PSRs are
compatible with the GC excess spectrum and $\mathcal{O}(1000)$ are
required to explain its intensity~\citep{Hooper:2010mq,
  2011JCAP...03..010A, Calore:2014oga,Cholis:2014lta}.  This
Galactic bulge pulsar population is hypothesized to be
distinct from the well-known ``disk'' population that follows the
Galactic spiral arms and from which we detect the mostly local known
sample of pulsars in radio and $\gamma$
rays~\citep{2005AJ....129.1993M,2013ApJS..208...17A}.  See, in
particular, Figure~2 of ~\citet{Calore:2015bsx} for an illustration
of the Galactic disk and bulge pulsar populations.  Finally, since
this PSR population would be distributed in the Galactic bulge its
spatial morphology could be consistent with that of the GC excess.

The large amount of data collected by the LAT after 7.5~years of
operation and the improvement in energy and spatial resolution brought
by Pass~8 enable a deeper search for PSRs in the Galactic bulge.  Such
a search is highly relevant to testing the potential PSR nature of the
GC excess.

Prospects for detecting radio pulsations from the bulge pulsar population
were studied by~\citet{Calore:2015bsx}, and the authors found that
existing radio pulsar surveys using the
Parkes~\citep{2010MNRAS.409..619K} and Green Bank~\citep{2013CQGra..30v4003S}
telescopes are not quite sensitive enough
to detect many pulsars from the bulge population.  On the other hand, large
area surveys using, e.g., MeerKAT and later
SKA~\citep{2015aska.confE..36K} should detect dozens to
hundreds of pulsars from the Galactic bulge.

In this paper we investigate the pulsar interpretation of the GC
excess deriving a new catalog of sources detected in the GC region and selecting among them PSR candidates using SED-based criteria.  
In Section~\ref{sec:datasel} we present the data selection
and the background models that we use. In Section~\ref{sec:sourcelist}
we describe our analysis pipeline and derive a list of sources in the
GC region. In Section~\ref{sec:systsource} we discuss potential 
systematic biases between the source lists and PSR candidate lists derived with two different IEMs.
In Section~\ref{sec:psrblazar} we study the SEDs of blazars
and PSRs detected by the LAT and introduce a criterion to select PSR
candidates from our list of sources. In Section~\ref{sec:lumi_dist} we
study the distribution of luminosities of known $\gamma$-ray PSRs.
Details of the data analysis pipeline
are provided in the appendices.

\section{Data selection and background models}
\label{sec:datasel}
The analysis presented in this paper uses 7.5~years of {\it Fermi}-LAT
data recorded between 2008 August 4 and 2016 February 4 ({\it Fermi}
mission elapsed time 239557418--476239414 s).  We apply the standard
data-quality selections\footnote[2]{See
  \url{http://fermi.gsfc.nasa.gov/ssc/data/analysis}.}.

Since we are interested in detecting the emission from point sources,
we select events belonging to the ``Pass 8 Source'' event class and
use the corresponding \texttt{P8R2\_SOURCE\_V6} instrument response
functions.  In order to reduce the contamination of $\gamma$ rays generated by 
cosmic-ray interactions in the upper atmosphere to negligible levels
we select events with a maximum zenith angle of
$90^{\circ}$. We use events in the energy range $E=[0.3,500]$~GeV.

We select data for a region of interest (ROI) that is a square of side
$40^{\circ}$ centered on the GC $(l,b)=(0^{\circ},0^{\circ})$, where
$l$ and $b$ are the Galactic longitude and latitude, since the GC
excess has an extension of approximately $20^{\circ}$.

We employ two different IEMs to estimate the systematic uncertainties
introduced by the choice of IEM.  The IEMs are brightest in the
Galactic disk where the density of interstellar gas and radiation
fields is greatest.  Additionally, isotropic emission, mainly due to
$\gamma$-ray emission from unresolved sources~\citep[see,
e.g.,][]{DiMauro:2015tfa} and residual contamination from interactions
of charged particles in the LAT misclassified as $\gamma$ rays, is
included in the model~\citep{2015ApJ...799...86A}.

The first IEM we use is the {\it gll\_iem\_v06.fits} template,
released with Pass~8 data~\citep{2016ApJS..223...26A}. The corresponding
isotropic component is the {\it iso\_P8R2\_SOURCE\_V6\_v06.txt}
template~\footnote[3]{For descriptions of these templates see
  \url{http://fermi.gsfc.nasa.gov/ssc/data/access/lat/BackgroundModels.html}.}.
These are routinely used for Pass~8 analyses and we refer to this
model as the official (Off.) model. 
The second IEM is the Sample model~\citep[][Section~2.2]{TheFermi-LAT:2017vmf}, from which we remove the GC excess component 
and add the {\it Fermi} Bubbles template at $|b|<10^{\circ}$~\citep[][Section~5.1.3]{TheFermi-LAT:2017vmf}.  
We refer to that model as the alternate (Alt.) model.


\section{Analysis pipeline and source list}
\label{sec:sourcelist}
We use the \fermipy Python package (version 00-11-00)\footnote[5]{See
  \url{http://fermipy.readthedocs.io/en/latest/}.} in conjunction
with standard LAT \texttt{ScienceTools}\footnote[6]{See
  \url{https://fermi.gsfc.nasa.gov/ssc/}.} (version 10-03-00)
to find and characterize point
sources for both IEMs.

To break the analysis into manageable portions we subdivide
the $40^{\circ}\times40^{\circ}$ ROI into 64 smaller
$8^{\circ}\times8^{\circ}$ ROIs with an overlap of $3^{\circ}$ between
adjacent ROIs.  Sources near the edge of an ROI are thus well
contained in an adjacent ROI.  Considering the entire
$40^{\circ}\times40^{\circ}$ ROI would imply several hundred free
parameters, making the analysis with the LAT \texttt{ScienceTools}
prohibitive.  In each ROI we bin the data with a pixel size of
$0.06^{\circ}$ and 8 energy bins per decade.  In general we analyze
each ROI separately; however, as discussed below, at certain points in
the analysis we merge information from the analyses of the different
ROIs.

The first step of the analysis is to find sources in each of the 64
ROIs.  For each ROI we construct an initial model consisting of the
IEM, the isotropic template and sources detected with $TS>49$ in the
{\it Fermi} LAT Third Source Catalog~\citep[3FGL,][]{2015ApJS..218...23A}.
This provides a reasonably good initial representation of the
$\gamma$-ray data in each ROI.  The procedure selects 116 3FGL sources that we
include in the $40^{\circ}\times40^{\circ}$ ROI.  As we will show
later in this section we recover the vast majority of the least significant 3FGL
sources (i.e., those with $TS$ values ranging from 25 to 49).

We then use \fermipy tools to refine the positions and the SED
parameters of 3FGL sources for the larger, Pass~8 data set that we use
here, as well as to find new sources in each ROI.  The details of this
procedure are described in Appendix~\ref{sec:fermipy}.  Since the ROIs
overlap slightly, as part of this procedure we remove duplicate
sources found in more than one ROI.  

\begin{figure}
	\centering
\includegraphics[width=1.03\columnwidth]{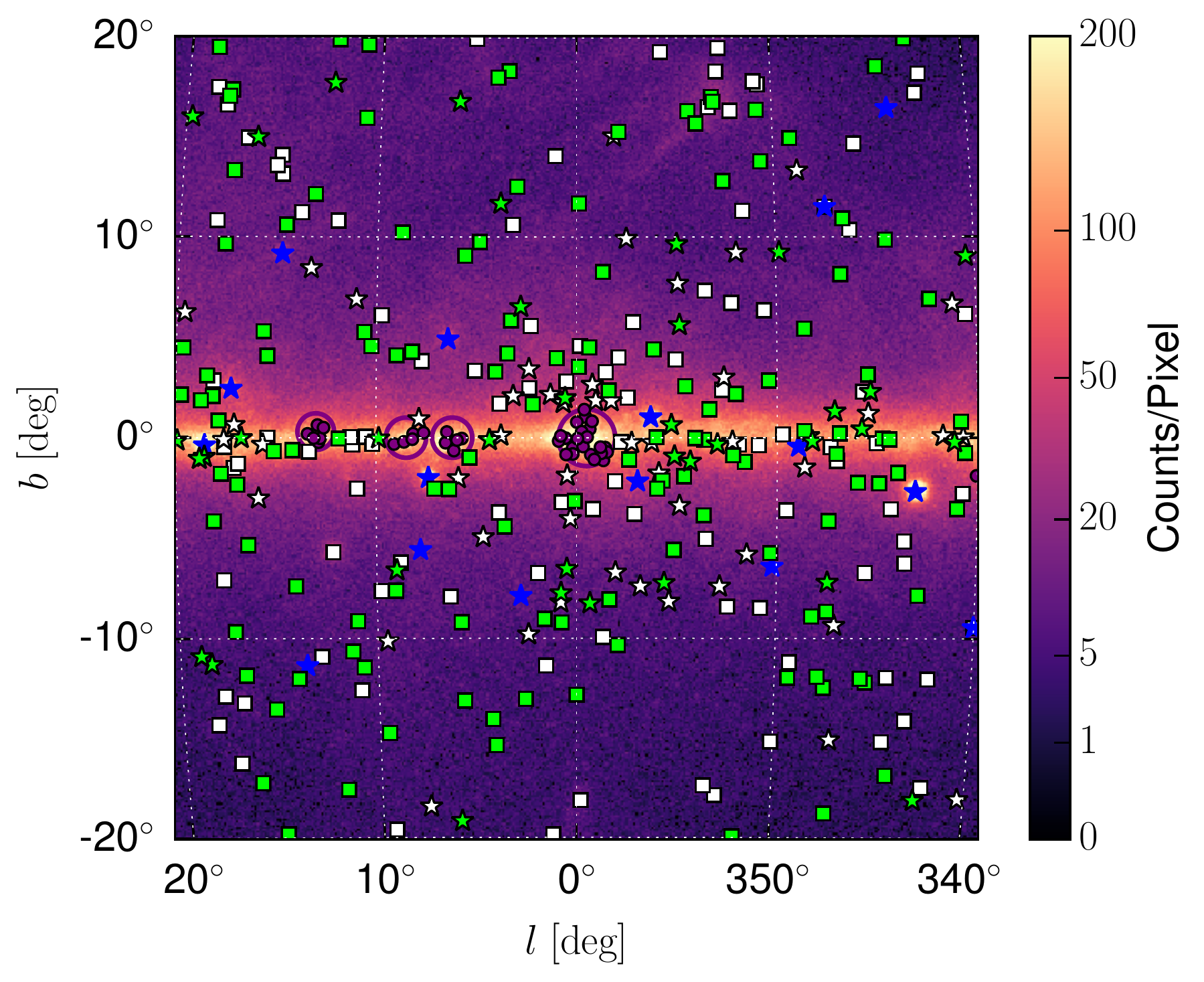}
\caption{Counts map of the $40^{\circ}\times 40^{\circ}$ ROI used in
  this analysis.  The map includes data for the range $[0.3,500]$ GeV.
  The map is in Hammer-Aitoff projection, centered on the GC and in Galactic
  coordinates.  The pixel size is $0.1^{\circ}$. The color scale shows
  the number of photons per pixel. Markers are shown at the positions
  of sources found in our analysis with the Off. IEM.  White markers show
  sources associated with a 3FGL source and green markers show new sources
  with no 3FGL counterpart.  Stars (squares) indicate sources that are
  (not) PSR-like and purple markers indicate sources belonging
  to a cluster, and the clusters are outlined with purple circles (see text for details). 
  Finally, blue stars show PSRs identified as or associated in the 3FGL.}
\label{fig:map} 
\end{figure}

We detect 374 (385) sources with $TS>25$ when using the Off. (Alt.)
IEM model.  Combining the list of detected sources with each IEM we detect 469 unique
sources of which 290 are found with both models.  The positions of
these sources are displayed in Figure~\ref{fig:map}, overlaid on a
counts map for the $40^\circ\times40^\circ$ ROI.  By comparison, the
3FGL catalog contains 202 sources in this region and 189 (182) of them
are found with our analysis with the Off. (Alt.)  IEM. The 1FIG,
which covered only the inner $15^{\circ}\times15^{\circ}$, contains 48 sources of
which we find 38 (41) when we employ in the analysis the
Off.~(Alt.) IEM.  We define associations with 3FGL and 1FIG sources
based on the relative positions and the $95\%$ localization uncertainty regions
reported in those catalogs and found in our analysis.  Specifically,
we require that the angular distances of sources in the 3FGL or 1FIG
from matching sources in our analysis be smaller than the sum in quadrature of
the $95\%$ containment angles in 3FGL or 1FIG and in our analysis.
The 3FGL and 1FIG sources that are not present in our lists either
have TS near the detection threshold (i.e., $25<TS<36$) or are
located within $0\fdg5$ of the GC.

The GC region is the brightest in the $\gamma$-ray sky and developing
a model of the interstellar emission in this region is very
challenging~\citep[see, e.g.,][]{Calore:2014xka,
  2016ApJ...819...44A, TheFermi-LAT:2017vmf}.  
Imperfections of our IEMs could manifest themselves
as dense concentrations of sources in regions where the IEMs particularly under-predict the diffuse intensity.  
To account for this, we
employed a source cluster-finding algorithm (described in
Appendix~\ref{sec:mst}) to identify such regions.  We
find a total of four clusters of sources with four or more sources
within $0\fdg6$ of at least one other source in the cluster.  These
clusters are located around the GC, in regions around the W28 and W30
supernova remnants and near 3FGL J1814.1$-$1734c, which is an
unassociated source in the 3FGL catalog.  (The `c' designation means
that it was flagged in that catalog as possibly an artifact.) These
clusters are shown in Figure~\ref{fig:map}.

We removed from further consideration here all sources identified as
belonging to clusters.

\section{Systematic biases in source-finding and
  PSR selection criteria efficiency from uncertainties in the IEMs}
\label{sec:systsource}

In this section we discuss potential systematic biases
  between the source lists and PSR candidate lists constructed using
  the ``Off.'' and ``Alt.'' IEMs.  We find that away from the Galactic
  plane ($|b| > 2.5^\circ$) these systematic uncertainties are
  similar in magnitude to the statistical uncertainties, and that along the
  plane the systematic differences between the two models are randomly
  enough distributed that they do not introduce significant biases to
  the analysis.  Finally, we argue that the effect of the
  uncertainties of the IEMs on the analysis of the Galactic bulge PSR
  population can be quantified reasonably by evaluating the difference
  in the results obtained using the two different IEMs.

\begin{figure*}[!ht]
  \centering
\includegraphics[width=0.995\columnwidth]{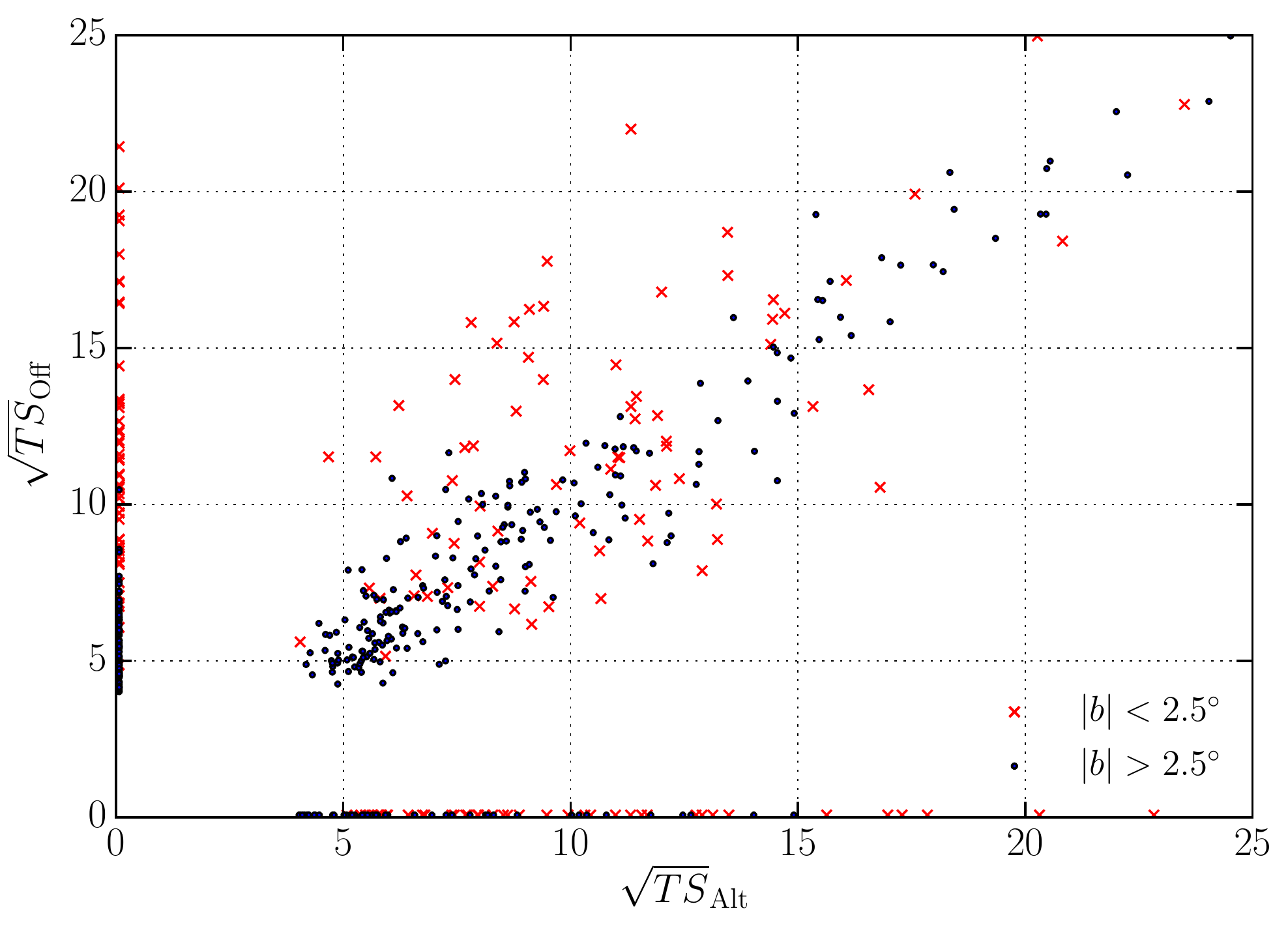}
\includegraphics[width=0.995\columnwidth]{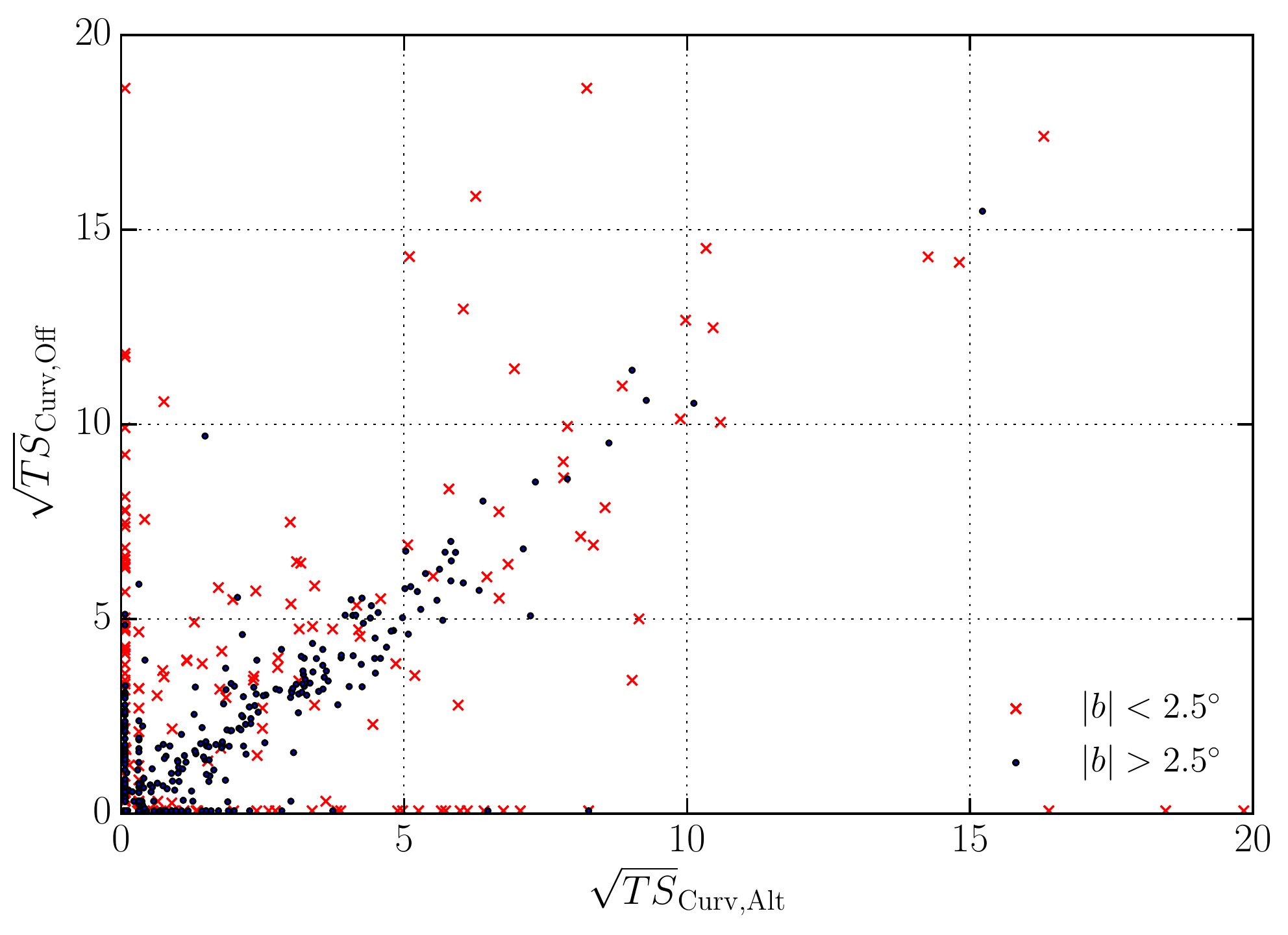}
\caption{Left: correlation between the significance ($\sqrt{TS}$)
  derived using the Off. and Alt. IEMs. The black (red) points are for
  sources with $|b| > 2.5^\circ$ ($|b| < 2.5^\circ$).  Sources that
  are detected with $TS > 16$ for only one of the two IEMs are
  assigned a $TS$ value near zero for the other IEM for plotting
  purposes.  Right: same, but for $\sqrt{TS_{\rm curv}}$; again,
  sources detected with both only one IEM have been a $TS$ value near
  zero for the other IEM.}  
\label{fig:ts_correl} 
\end{figure*}

Following the method we used to associate sources with the 
  3FGL and 1FIG lists, we match sources between the lists produced with
  the Off. and Alt. IEMs if the separation between two positions is
  less than the sum in quadrature of the 95\% containment radii of the
  source localization.  In the left panel of Figure~\ref{fig:ts_correl} we show the significance (i.e.,
  $\sqrt{TS}$) as measured with the two IEMs,
  considering all source candidates with $TS > 16$ in either IEM.  We
  note the high degree of correlation between the significance
  found with the two different IEMs; the absolute scatter (i.e., the
  RMS of the difference) of the sources found using both IEMs 
  is $1.7\sigma$ ($4.6\sigma$)  for sources with $|b| > 2.5^\circ$
  ($|b| < 2.5^\circ$).   Furthermore, sources were detected with
  slightly larger significance on average with the Alt. IEM; the
  difference was $0.2\sigma$ ($1.8\sigma$)  for sources away from
  (along) the plane.

Many of the sources are faint and are detected with TS near
  the threshold of 25.  On the other hand, the requirement $TS_{\rm
    curv} > 9$ in the PSR selection criteria effectively sets a
  somewhat higher flux threshold, as it is difficult to measure
  spectral curvature for a faint source near the detection threshold.
  Therefore it is equally important to consider the effects of
  fluctuations near threshold on $TS_{\rm curv}$.  The
  $\sqrt{TS_{\rm curv}}$ values found using the two IEMs are
  shown in the right-hand panel of Figure~\ref{fig:ts_correl}.  The
  absolute scatter of $\sqrt{TS_{\rm curv}}$ is 1.0 (3.0) for
  sources with $|b| > 2.5^\circ$ ($|b| < 2.5^\circ$).

\begin{figure}[!ht]
  \centering
\includegraphics[width=1.005\columnwidth]{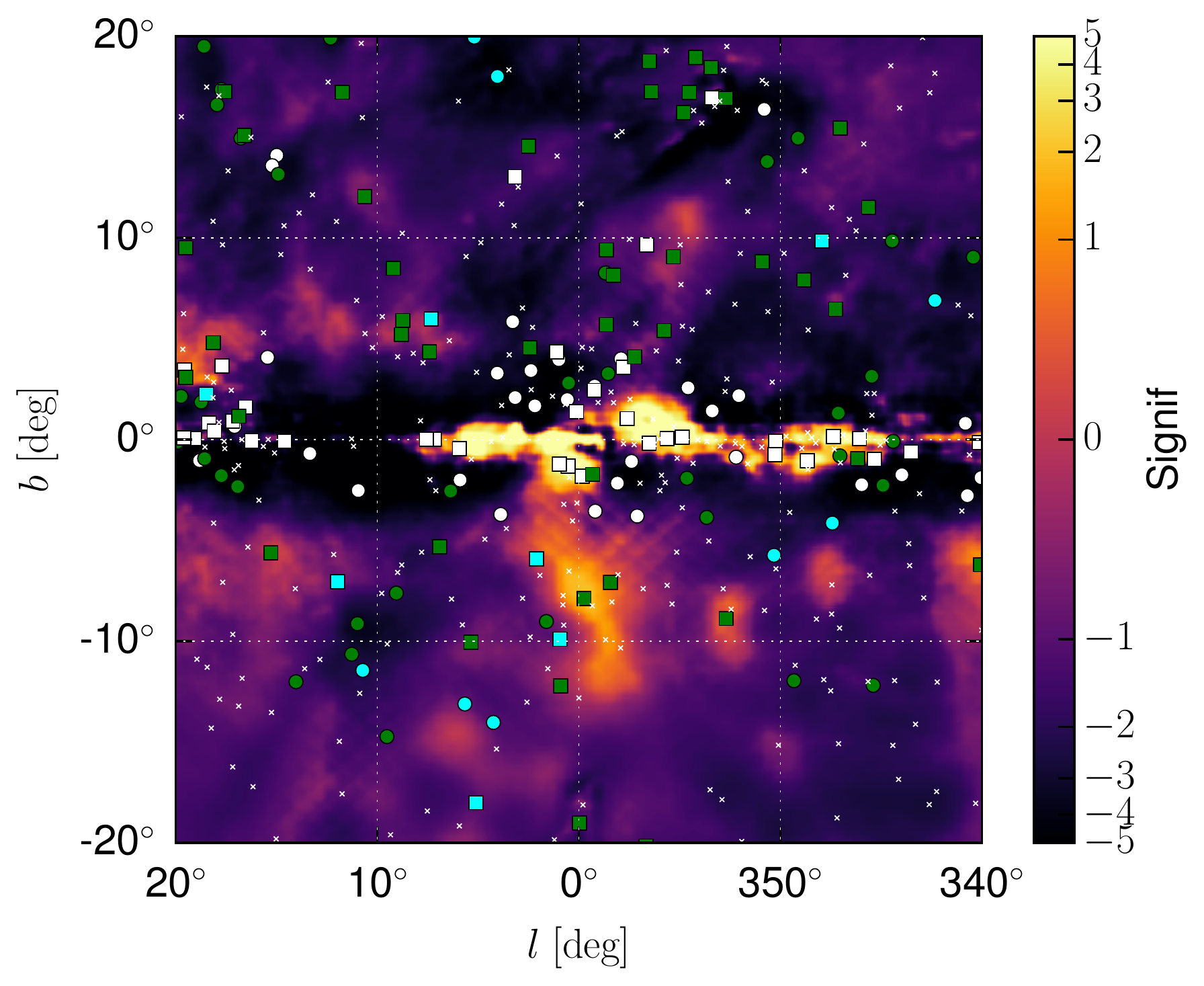}
\caption{The color scale shows the statistical significance of the
  difference between the two IEMs integrated over the range
  $E = [0.3,500]$~GeV.  Sources found with both IEMs are indicated
  with 'x' markers.  Square (circle) markers show sources found only
  with the Off. (Alt.) IEM.  The cyan markers show sources that have a
  corresponding sub-threshold candidate with $16 \le TS < 25$ using the other IEM.  Green
  markers show low significance sources with $25 \le TS <
  49$;  white circles and squares show high signficance sources with $TS \ge 49$.}
\label{fig:systmap} 
\end{figure}

We also explicitly study the differences between the two
  IEMs and how those correlate with differences in the sources 
  of the respective lists.  In Figure~\ref{fig:systmap} we compare locations
  of sources found with only one of the two IEMs with the estimated
  statistical significance of the difference between the two.
  Specifically we used the {\texttt{gtmodel}} tool to produce the
  expected counts from diffuse IEMs (including the isotropic
  component) and computed the difference divided by the square-root of
  the mean of the two IEMs.  The white boxes and circles show sources that are
  detected with $TS > 49$ using one IEM, but that are not detected
  with the other, and thus primarily attributable to differences in the
  two IEMs.  The large majority of such cases occur near the Galactic
  plane, which is unsurprising as it is also the region where the
  difference between the two IEMs is the most significant.  On the
  other hand, the differences between the two IEMs are less
  statistically significant (generally between $\pm 1 \sigma$) away
  from the plane, and most of the sources found with only one of the
  two IEMs are either near threshold (i.e, $25 \le TS < 49$, green markers)
  or are seen as sub-threshold source candidates ($16 \le TS <  25$, cyan
  markers) with the other IEM.

From these studies we conclude that away from the Galactic
  plane ($|b| > 2.5^\circ$) the differences between the two 
  lists are largely attributable to the combination of thresholding
  and variations between the IEMs that are sub-dominant to the statistical
  variations of the data.  Furthermore, $TS = 25$ for four degrees of
  freedom (source position, spectral index and normalization)
  corresponds to a detection at the level of 4.1 standard deviations,
  and the systematic uncertainties in the source significance are
  smaller than the statistical uncertainties.  Therefore, even
  considering the systematic uncertainties attributable to the IEMs, the
  pre-trials significance of these source candidates is close to 3 standard
  deviations.

To estimate how much of the difference between the two lists
  can be attributed to the combination of small systematic variations
  with thresholding effects, we simulated the effect of the
  significance variations attributable to the differences between the two IEMs by adjusting the
  $\sqrt{TS}$ of each source by a random number drawn from a Gaussian
  distribution with mean and width given by $\mu = 0$, $\sigma=1.7$
  ($\sigma=2.6)$ for sources away from (along) the Galactic plane and
  testing if the adjusted $TS$ value was above the $TS = 25$ detection
  threshold.  In this simulation we found that 79\% (58\%) of the
  sources found with either IEM would be expected to be found with
  both IEMs, for sources away from (along) the Galactic plane.  
  In the actual lists, these numbers were 64\% and 56\%.  Since
  the scatter away from the plane is small compared to the $5.0\sigma$
  detection threshold, we believe that ratio between the 79\%
  overlap found in our toy simulation and the 64\% overlap found in
  the two cases sets a lower bound that $64/76 = 0.81$ of the sources
  found away from the plane are with either IEM are in fact real. 
  Along the plane, on the other hand, the scatter is comparable in
  magnitude to the detection threshold.  There we believe that the 
  56\% overlap of sources found with both IEMs the lower bound
  on the fraction of real sources.

We then studied individual ROIs near the Galactic plane 
  to better understand the interplay between the implementation details
  of our data analysis pipeline and the difference in the IEMs.  Specifically, 
  since we fit the normalization of the Galactic diffuse emission
  in each of the ROIs, and iteratively add point sources 
  in the largest positive residuals, these steps give the resulting
  models some freedom to compensate for inaccuracies in the IEMs.

\begin{figure}[!ht]
  \centering
\includegraphics[width=1.005\columnwidth]{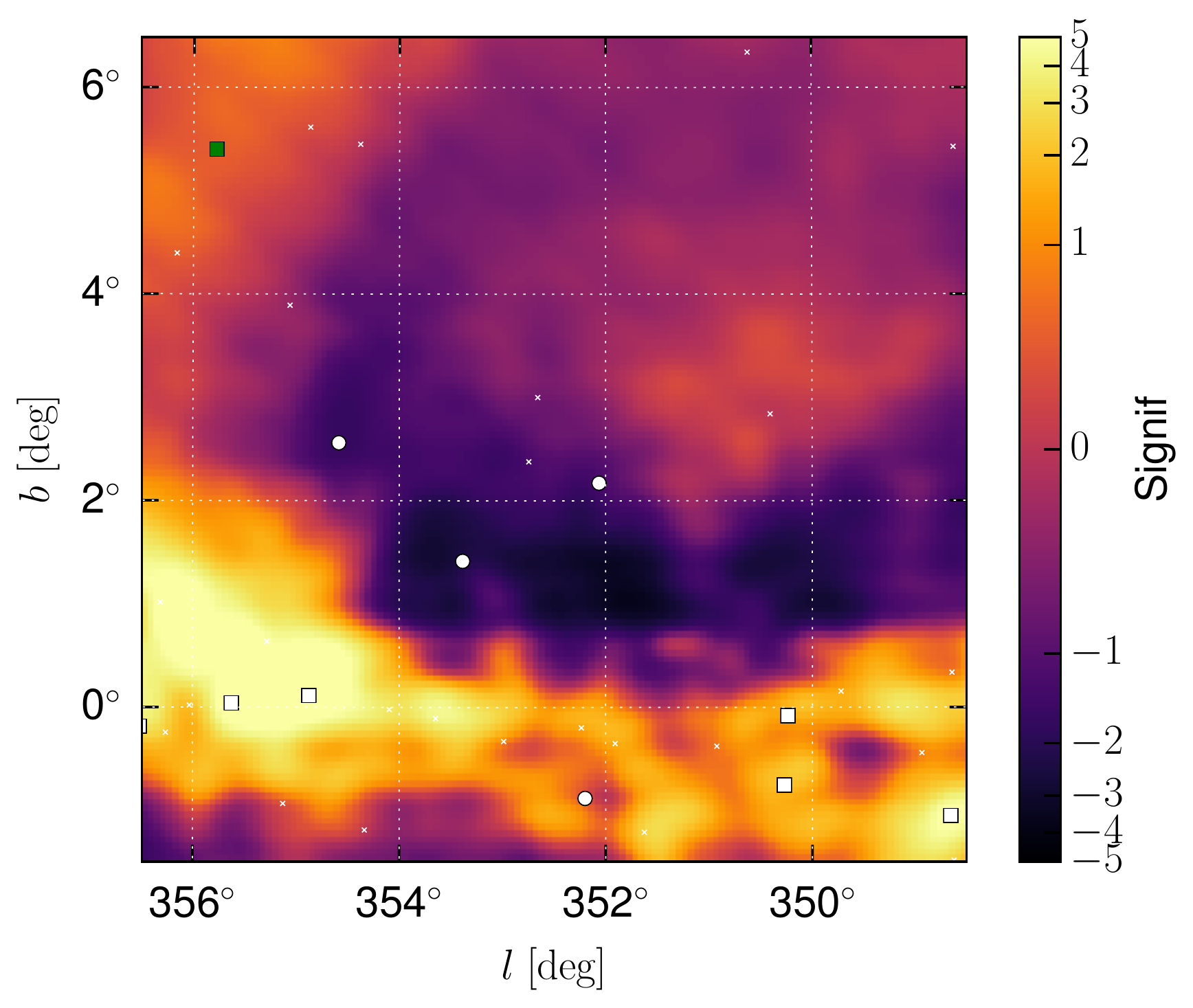}
\caption{Same as Figure~\ref{fig:systmap}, but for one of the $8^{\circ}\times8^{\circ}$
  ROIs used in the analysis pipeline, and for the complete model map at the
  end of the analysis pipeline, using the fitted values of the IEM normalization.}
\label{fig:systmap_roi43} 
\end{figure}

Figure~\ref{fig:systmap_roi43} is similar to Figure~\ref{fig:systmap}, 
  but was produced for a single one of our $8^{\circ}\times8^{\circ}$ ROIs, and 
  using the fitted values of the IEM normalization.

\begin{figure*}[!ht]
  \centering
\includegraphics[width=0.995\columnwidth]{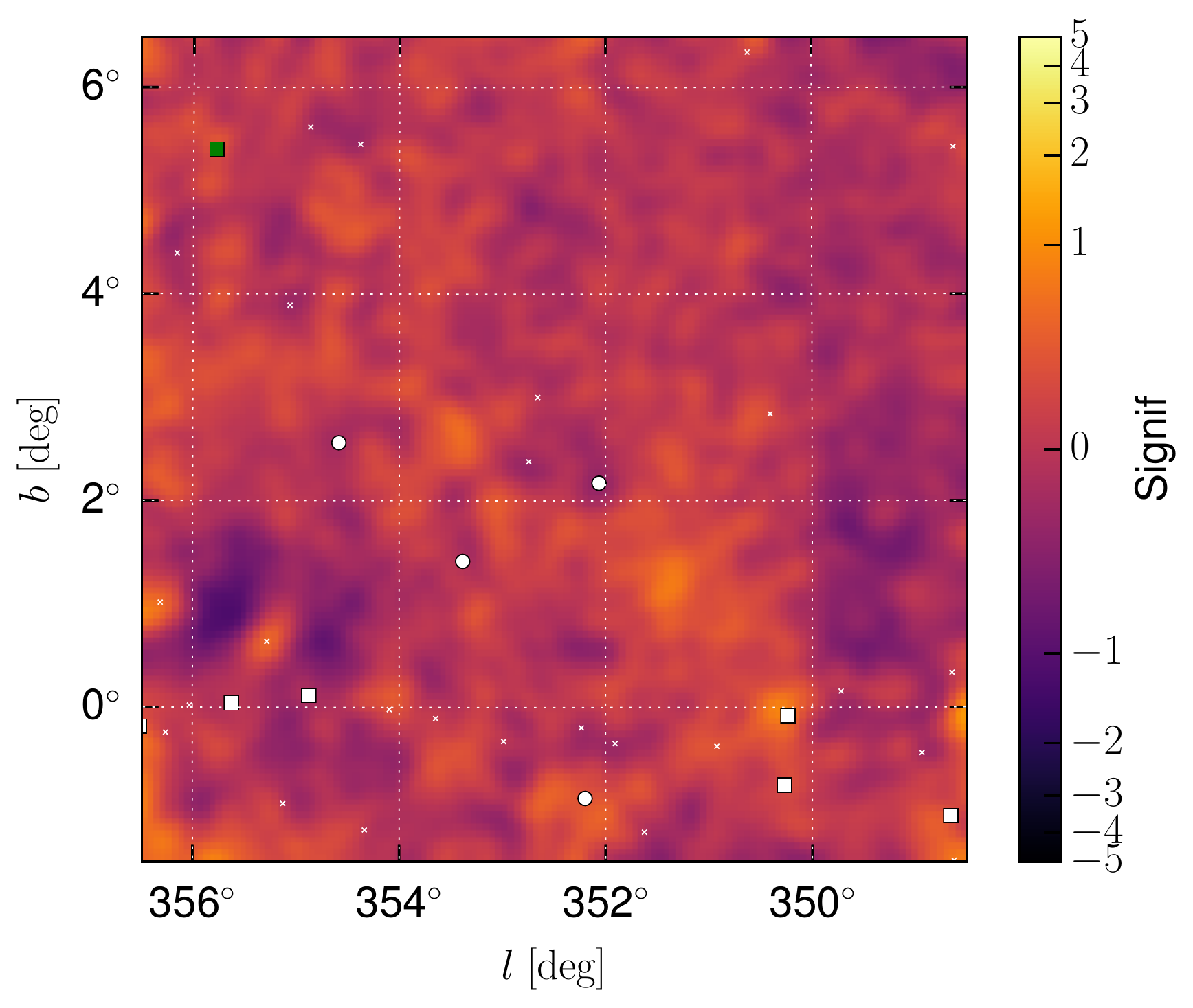}
\includegraphics[width=0.995\columnwidth]{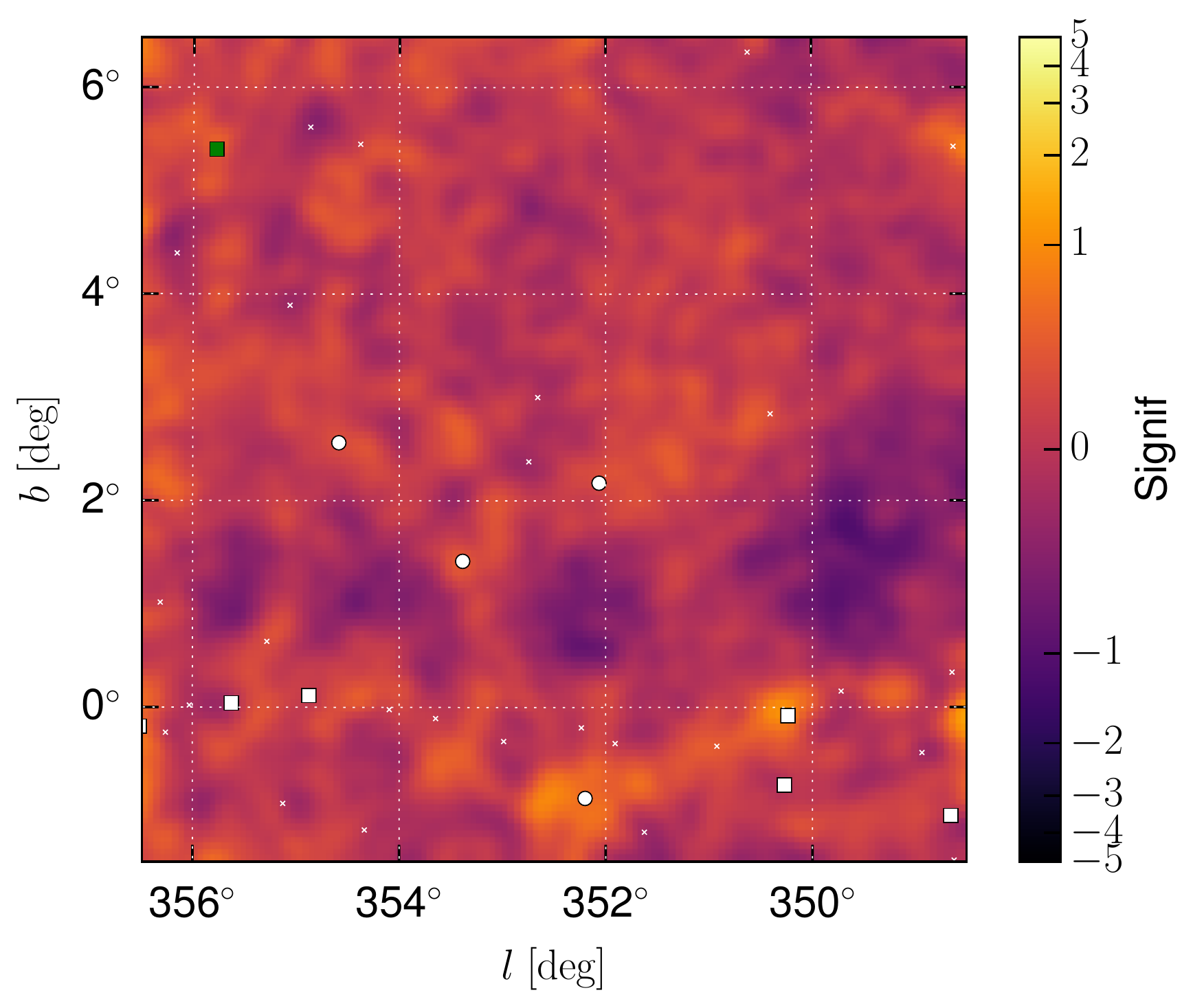}
\caption{Estimated significance of residuals, $({\rm data} - {\rm model})/\sqrt{\rm model}$, 
  for the Off. (left) and Alt. (right).   The markers have the 
  same meanings as for Figures~\ref{fig:systmap} and \ref{fig:systmap_roi43}.
  Gaussian smoothing with a width of $\sigma=0.18^\circ$ has been applied.}
\label{fig:signif_resid} 
\end{figure*}

Figure~\ref{fig:signif_resid} shows the significance of the residuals 
  for the same ROI as computed with the two different IEMs.  Overall,  
  some correlation is evident between the residuals in the two maps,
  but there are no regions larger than a few degrees across where both models over-predict the data.

\begin{figure}[!ht]
  \centering
\includegraphics[width=0.905\columnwidth]{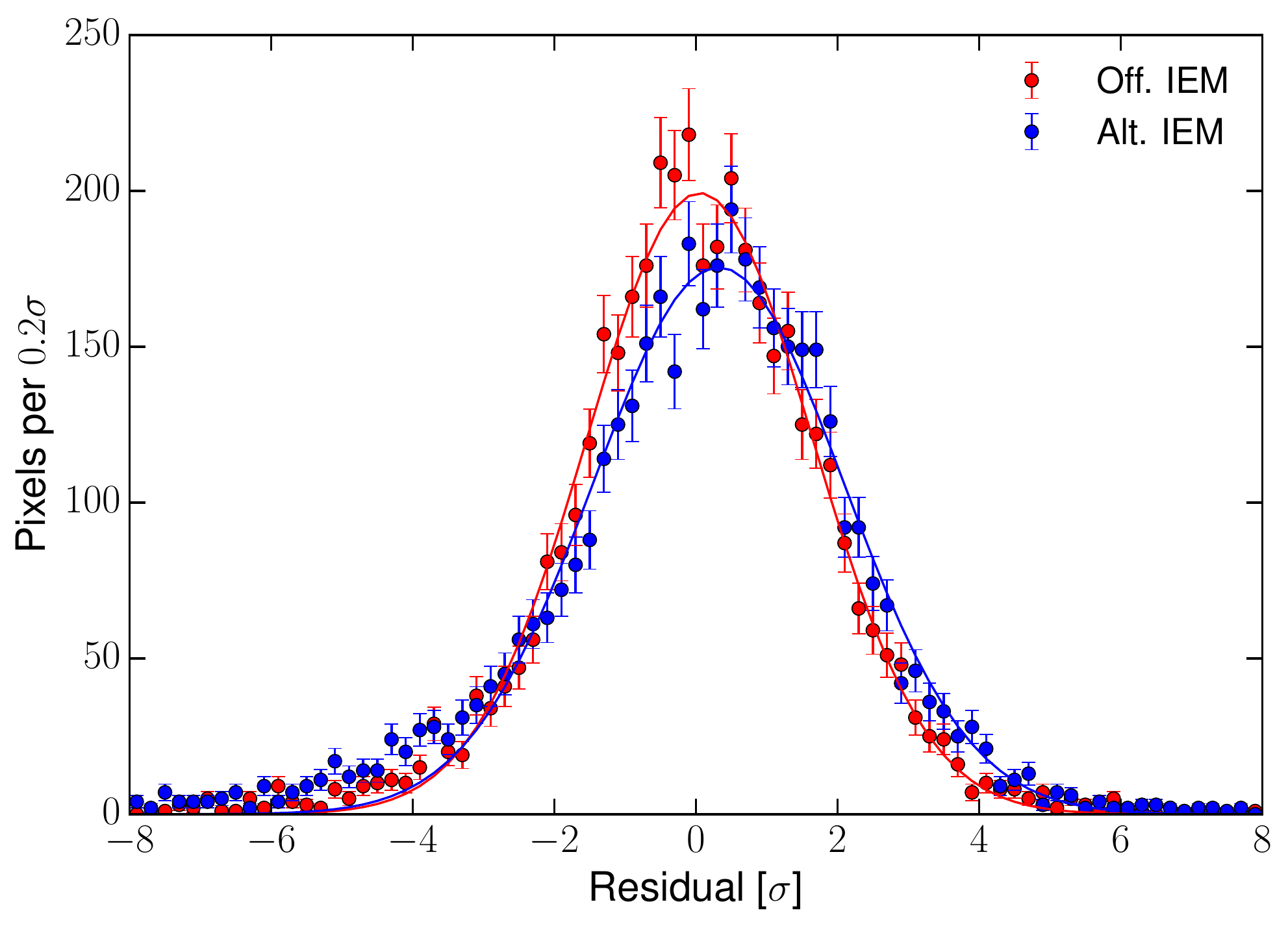}
\caption{Distribution of estimated significance of residuals for 
  $0.48^\circ \times 0.48^\circ$ spatial bins in the 16 ROIs along the Galactic plane.  The
  solid lines show Gaussian fits to the two distributions.  The best-fit
  mean and width are $\mu = 0.1, \sigma = 1.6$ ($\mu = 0.3, \sigma = 1.8$) for the
  Off. (Alt.) IEM.}
\label{fig:resid_hist} 
\end{figure}

To quantify the finding that the two IEMs do not both over-predict
  the data in the same regions, we have evaluated histograms of the estimated
  significance of the residuals, $({\rm data} - {\rm model})/\sqrt{\rm model}$, for 
  $0.48^\circ \times 0.48^\circ$ spatial bins (we simply rebinned the
  model and data maps combining $8\times8$ pixels) for both IEMs for all 16 ROIs along the
  Galactic plane (Fig.~\ref{fig:resid_hist}).

These distributions are fairly similar for the two IEMs. 
  The distribution for the Off. IEM is narrower and more symmetric and
  is well described by a Gaussian with $\mu=0.1$ and $\sigma=1.6$.  
  On the other hand, for the Alt. IEM, the distribution is wider, and
  has a non-Gaussian tail extending to around $5 \sigma$.  This
  indicates that there are in fact regions where the Alt. IEM
  over-predicts the data, but that they are moderately-sized.
  Since we obtain similar results with the Off. and Alt IEMs,  we
  believe that the dominant effects of the uncertainties of the IEMs
  is to widen the effective threshold.   Since there are more sources
  just below threshold that just above threshold, we expect that this
  will cause us to detect more sources that we would see without
  the effect of the IEMs.

From this we conclude that any biases in the source lists due
  to the IEMs are likely to be slight overestimates of the number of
  sources.  These overestimates would be somewhat larger along the
  Galactic plane than away from it.

\begin{figure}[!ht]
  \centering
\includegraphics[width=0.95\columnwidth]{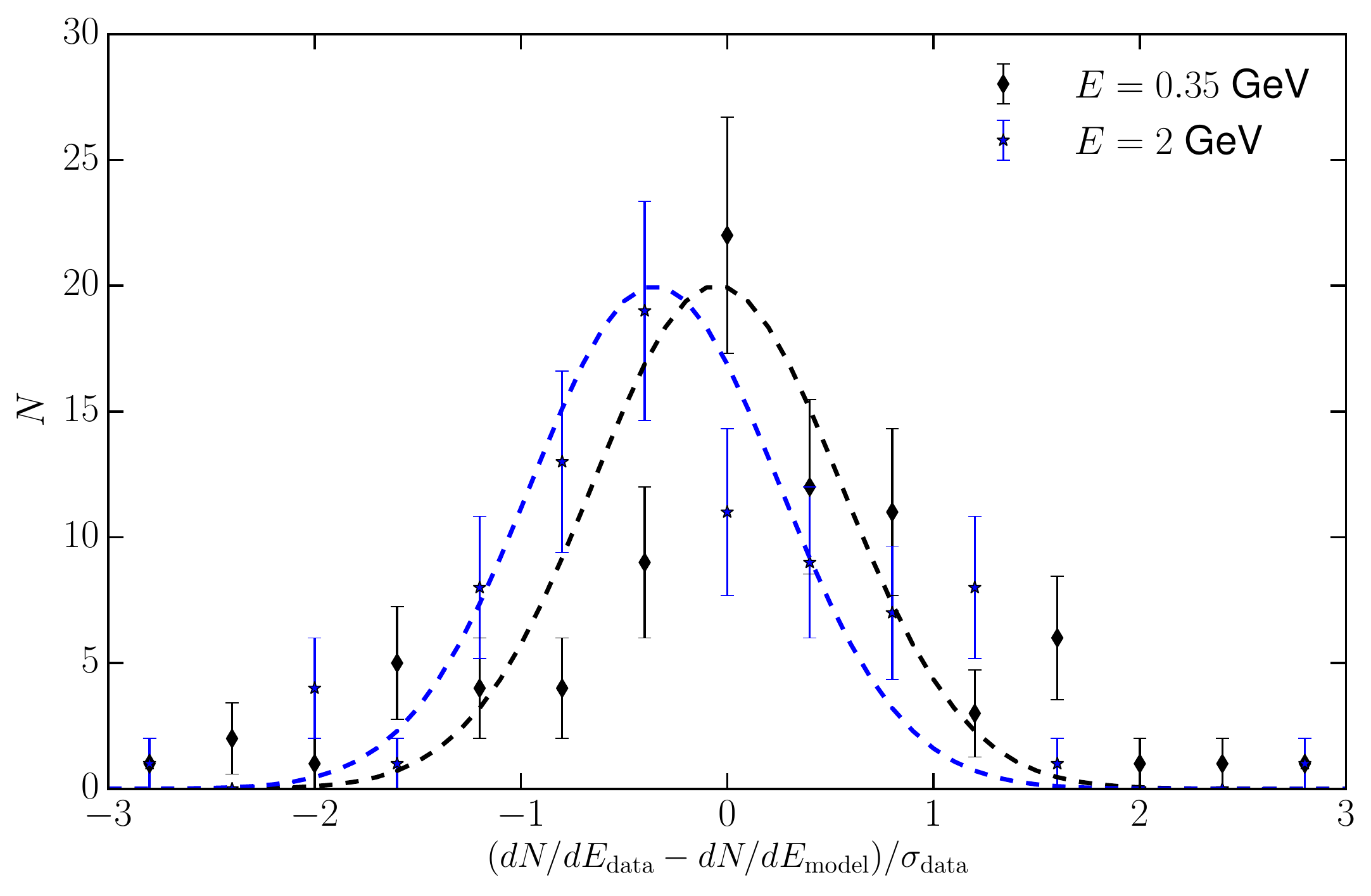}
\caption{Distribution of the significance of the residual for the
  energy flux in a single bin with respect to fitted spectral model
for the PSR candidates found with the Off. IEM analysis.}
  This is shown for the lowest-energy bin (black), and for an energy bin
  near the peak of the statistical power of the analysis ($\sim
  2$~GeV, blue).  The solid curves are best-fit Gaussian distributions
  to the two histograms; the best-fit parameters are given in the text.
\label{fig:signif_bin} 
\end{figure}

We have also studied the possibility that biases in the IEMs
  could affect the spectral parameters of the sources, causing either
  true pulsars to be rejected, or increasing the number of non-pulsars
  mistakenly selected as PSR candidates.  We found that the
  correlation between the spectral parameters and the background model
  is modest: the correlation coefficient between the normalization of
  the IEM and the $\Gamma$ and $E_{\rm cut}$ spectral parameters for
  the new sources ranges from $-0.25$ to $0.10$.  Given the very large
  statistics along the Galactic plane, small fractional biases in the
  IEM could cause marked biases in the spectral parameters.
  Therefore, we have looked at the agreement between the energy fluxes
  in the low-energy bins and the spectral models and found no evidence
  of significant biases caused by the IEMs.  In
  Figure~\ref{fig:signif_bin} we present a histogram of the residual
  of the single- energy bin fluxes with respect to the spectral
  models, for two different energy bins.  For the lowest energy bin
  the scaled residuals are very nearly normally distributed
  ($\mu = 0.05$, $\sigma = 1.0$); for the energy bin near 2~GeV model
  slightly overestimates the data ($\mu = -0.2$, $\sigma = 1.0$).
  Overall we believe that these results indicate that the spectra are
  not significantly biased by any errors in the IEMs.

From these studies we conclude that the systematic variations
  near the plane appear to be distributed in such a manner that no
  substantial regions have both IEMs either under- or over-estimate
  the data.  We also conclude that the effect on the measured spectral
  parameters is small.  Therefore, we believe that the systematic
  uncertainties in the properties of the PSR population are well
  described by the differences obtained with the two IEMs, or would
  tend to bias the fitting procedure to find lower numbers of PSRs in
  the Galactic bulge.

\section{SED of pulsars and blazars} 
\label{sec:psrblazar}

\begin{figure*}
	\centering
\includegraphics[width=1.03\columnwidth]{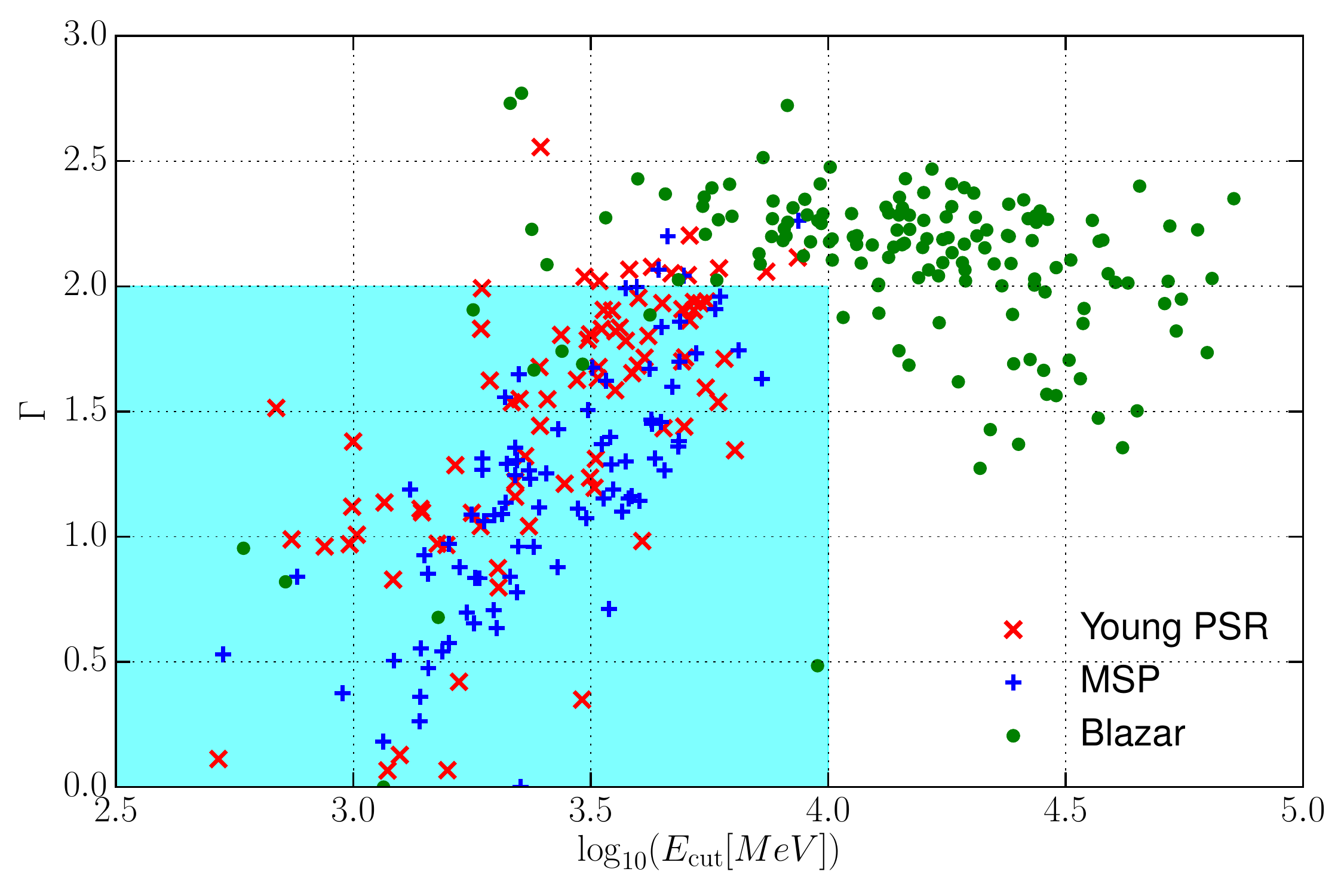}
\includegraphics[width=1.03\columnwidth]{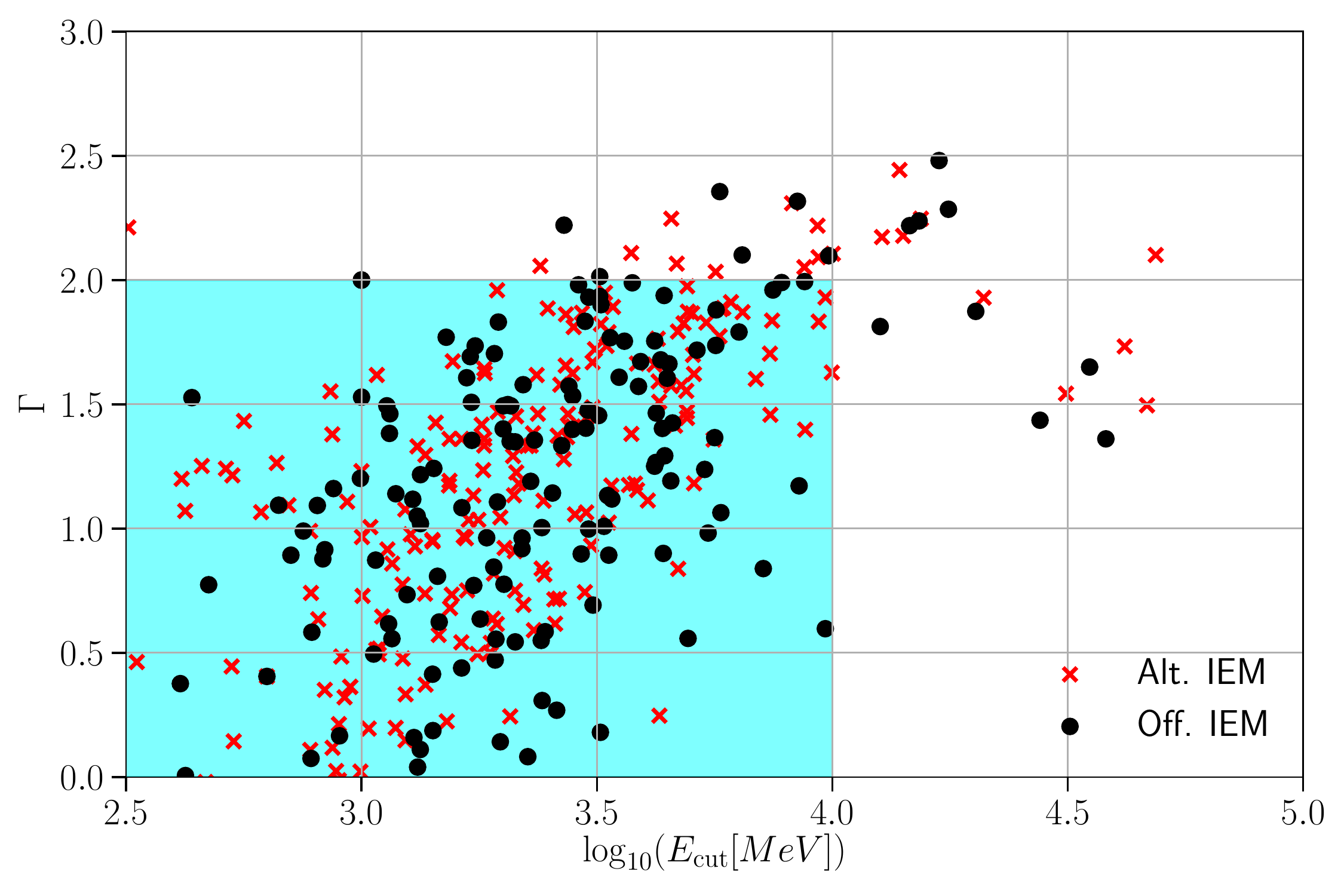}
\caption{Left: photon index $\Gamma$ and energy cutoff
  $E_{\rm{cut}}[\rm{MeV}]$ of PSRs and blazars detected 
  in our analysis with $TS^{\rm{PLE}}_{\rm{curv}}>9$. MSPs are shown
  as blue plus signs and young PSRs as red crosses. Here we are showing blazars in the
  3FGL~\citep{2015ApJS..218...23A} catalog with curvature significance as in
  the 3FGL ($\texttt{Signif\_Curve}$) larger than 3 (green circles).
  Right: same as in the left panel but applied to sources in our $40^{\circ}\times40^{\circ}$ ROI
  detected with $TS^{\rm{PLE}}_{\rm{curv}}>9$ in our analysis with the
  Off. IEM (black circles) and Alt. IEM (red crosses).}
\label{fig:psrblazar} 
\end{figure*}

In {\it Fermi}-LAT catalogs, blazars are the most numerous source
population.  Blazars are classified as BL Lacertae (BL Lacs) or Flat
Spectrum Radio Quasars (FSRQs) depending on the presence of strong
emission optical lines.  In the 3FGL $95\%$ of BL Lac and $85\%$ of
FSRQ spectra are modeled with a power-law (PL) function while blazars
with a significant spectral curvature (only about 10\% of the entire
population) are modeled with a log-parabola (LP)\footnote[7]{See
  \url{http://fermi.gsfc.nasa.gov/ssc/data/analysis/scitools/source\_models.html}
  for a description of the spectral models implemented in the LAT
  \texttt{ScienceTools}.}  
\citep[see, e.g.,][for a characterization of the {\it Fermi}
  blazar
  population]{2012ApJ...751..108A,DiMauro:2013zfa, Ackermann:2015yfk,Ghisellini:2017ico}. On
the other hand, a power law with exponential cutoff (PLE) at a few GeV
is the preferred model for pulsars~\citep{2013ApJS..208...17A}.  Of
the 167 PSRs reported in the 3FGL (143 PSRs identified by pulsations
and 24 sources spatially associated with radio pulsars) 115 have
spectral fits parametrized with a PLE because they have a significant
spectral curvature \citep[see, e.g.,][for a characterization
  of the $\gamma$-ray and radio pulsar
  population]{2004IAUS..218..105L,Calore:2014oga}.  The functional
definitions of the PL, LP, and PLE spectra are given
in~\citet{2015ApJS..218...23A}.

As described above, spectral shape is a promising observable to
separate PSRs from blazars.  We fit the spectrum of each source
in the ROI and derive the likelihood values for both the PL
($\mathcal{L}_{\rm{PL}}$) and PLE ($\mathcal{L}_{\rm{PLE}}$) spectra.
We introduce for each source in our analysis the TS for a curved
spectrum as:
$TS^{\rm{PLE}}_{\rm{curv}}=2\cdot(\log{\mathcal{L}_{\rm{PLE}}} -
\log{\mathcal{L}_{\rm{PL}}})$. This parameter quantifies the
preference to model an SED with a PLE with respect to a PL.

We perform the same analysis on known PSRs and blazars to study the
distribution of spectra of these two populations and develop criteria
to select PSR candidates.  We use the public list of $\gamma$-ray PSRs
with 210 sources\footnote[8]{See
  \url{https://confluence.slac.stanford.edu/display/GLAMCOG/Public+List+of+LAT-Detected+Gamma-Ray+Pulsars}.}
and the sub-sample of sources identified with or associated with
blazars in the 3FGL catalog that have significant spectral curvature.
Our blazar sub-sample includes all 3FGL blazars that have
\texttt{Signif\_Curve} greater than 3, where \texttt{Signif\_Curve} is
the significance in standard deviations of the likelihood improvement
between PL and LP spectra.  We use this sub-sample to study those
blazars most likely to be incorrectly flagged as PSR candidates.  This
reduced sample of blazars contains 218 objects.

Our definition of $TS^{\rm{PLE}}_{\rm{curv}}$ is slightly different
from the 3FGL \texttt{Signif\_Curve} parameter ($\sigma_{\rm{curv}}$)
in that $TS^{\rm{PLE}}_{\rm{curv}}$ is defined as the likelihood
improvement for a PLE spectrum with respect to the PL spectrum.
Furthermore, the 3FGL catalog analysis was based on only 4~years of
LAT data.  Therefore we used \fermipy to re-analyze
$10^{\circ}\times10^{\circ}$ ROIs centered around each source in this
sample of 210 $\gamma$-ray PSRs and 218 blazars.  From this
re-analysis we derived $TS^{\rm{PLE}}_{\rm{curv}}$, the photon index
($\Gamma$) and the energy cutoff ($E_{\rm{cut}}$) for the PLE
spectrum.  

Of the 210 PSRs, 172 (169) were found to have
$TS^{\rm{PLE}}_{\rm{curv}}>9$.  The average and standard deviation of
their photon indices and energy cutoffs were
$\Gamma=1.33\pm0.54 (1.30\pm0.54)$
and$\log_{10}(E_{\rm{cut}}[\rm{MeV}])=3.43\pm0.24 (3.40\pm0.24)$ when
we employed the Off.~(Alt.)~IEM.

In Table~\ref{tab:psrcandspectra} we
report the average photon index and cutoff energy for young PSRs
(rotational period $P$ greater than 30 ms) and millisecond PSRs
(MSPs). The energy cutoff parameter is consistent between young PSRs
and MSPs while the average photon index of MSPs is slightly harder.

In the sample of 218 blazars with \texttt{Signif\_Curve} $> 3$, 153 have
$TS^{\rm{PLE}}_{\rm{curv}}>9$.  In the left-hand panel of
Figure~\ref{fig:psrblazar} we show $\Gamma$ and
$\log_{10}(E_{\rm{cut}})$ for PSRs and blazars detected with
$TS^{\rm{PLE}}_{\rm{curv}}>9$. The two populations are well separated
in the plotted SED parameters. Taking $\Gamma<2.0$ and
$E_{\rm{cut}} < 10$~GeV as selection criteria (shown
in cyan in the figure) only 12 blazars, $7\%$ of our blazar sample and
less than $1\%$ of the entire 3FGL blazar population, are incorrectly
flagged as PSR candidates.  Clearly, these selection criteria are
effective for distinguishing the PSRs from blazars.  Additional studies
of the efficiency and false-positive rate of these selection criteria
using simulated data are described in Appendix~\ref{sec:testcurv}.  

We apply the PSR candidate selection criteria to our source lists.  In
the list derived with the Off. (Alt.) IEM we find 86 (115) PSR
candidates.  If we require that the source is selected with both IEMs
we find 66 PSR candidates.  In the right-hand panel of
Figure~\ref{fig:psrblazar} we show the $\Gamma$ and
$\log_{10}(E_{\rm{cut}})$ for all sources detected with
$TS^{\rm{PLE}}_{\rm{curv}}>9$ for the analysis with the Off.  IEM.
The average SED parameters for PLE are shown in
Table~\ref{tab:psrcandspectra}.  For PSR-like sources detected with
both IEMs the photon index ($1.02\pm 0.52$) is harder with respect to
known PSRs (see fifth row in Table~\ref{tab:psrcandspectra}).  This is
due to observational biases for the detection of PSRs in direction of
the inner part of our Galaxy.  We will show this in
Section~\ref{sec:bulge_dist}.  We also calculate the integrated energy
flux ($S = \int^{E_{\rm{max}}}_{E_{\rm{min}}} E dN/dE dE$) over the
range from $E_{\rm{min}}=300$ MeV to $E_{\rm{max}}=500$ GeV.


The full list of sources detected with $TS>25$ is provided as a FITS file and
described in detail in Appendix~\ref{sec:app2fig}.  We designate the sources with the prefix `2FIG'
designation; however we emphasize that many of the fainter sources in the
list are detected only with one of the two IEMs we used.  

Globular clusters are gravitationally bound concentrations of
ten thousand to one million stars and are the most ancient
constituents of our Milky Way Galaxy.  They are known to contain
many pulsars.  Among the detected sources we have
11 globular clusters already identified in the 3FGL and among those,
6 satisfy the PSR-like criteria.

\begin{table}[t]
\center
\begin{tabular}{crcc}
IEM & $N_{\rm{PSR}}$        & $\Gamma$       & $\log_{10}(E_{\rm{cut}} [\rm{MeV}])$   \\
\hline
Off.	&	86 &	$1.03\pm0.52$ &	$3.28\pm0.33$ \\
Alt.	&	115 &	$1.05\pm0.50$ &	$3.27\pm0.31$ \\
Alt. $\cap$ Off. (Off.) &	66 &	$1.02\pm0.52$ &	$3.27\pm0.32$ \\
Alt. $\cap$ Off. (Alt.) & 	66	& $1.01\pm0.51$ &	$3.26\pm0.30$ \\
Known PSRs (Off.) &	172	& $1.33\pm0.54$ &	$3.43\pm0.24$ \\
Young PSRs (Off.) &	86	& $1.46\pm0.53$ &	$3.44\pm0.26$ \\
MSPs(Off.) &	86	& $1.20\pm0.50$ &	$3.42\pm0.23$ \\
\end{tabular}
\caption{Spectral parameters of PSR candidates 
  compared with known PSRs.}
\tablecomments{Mean values and standard deviations of $\Gamma$ and 
  $\log_{10}(E_{\rm{cut}} [\rm{MeV}])$ for PSR candidates 
  compared with known $\gamma$-ray PSRs.  The first two 
  rows are found using different IEMs (Off. first and Alt.
  second row). The third (fourth) row is for the PSR candidates detected with 
  both IEMs, computed with the parameters derived in the analysis with the Off. (Alt.) IEM.
  The last three rows list the parameters for all $\gamma$-ray PSRs, young PSRs and 
  MSPs detected with $TS^{\rm{PLE}}_{\rm{curv}}>9$.}
\label{tab:psrcandspectra}
\end{table}


\section{Luminosity Distribution of PSRs}
\label{sec:lumi_dist}

Of the fraction of the 210 identified $\gamma$-ray PSRs for
which we have good distance estimates (roughly 100 have
  reported fractional uncertaintes smaller that 25\%), the large majority are located
within 4~kpc of the Solar System~\citep[see, e.g., Figure 3
of][]{2013ApJS..208...17A}.  They are thus ``local'' and belong to the
Galactic disk population.  A pulsar interpretation of the GC excess
requires a Galactic bulge population of PSRs.  (Throughout this paper
we adopt 8.5 kpc as the distance to the GC.)  However the known
Galactic disk PSR population is a strong foreground to the putative
Galactic bulge population.  In this section and the next we describe
simulations of the Galactic disk and bulge populations that are based
on the morphology and energy spectrum of the GC excess and the
characteristics of 3FGL PSRs.  We then use these simulations to
estimate the number of sources in these populations that would be
needed to match both the observed GC excess and the numbers and
properties of the detected sources in the $40^\circ \times 40^\circ$
ROI.

To perform these simulations, we use {\tt Fermipy} as explained in
Appendix~\ref{sec:appsim} to generate simulated data sets of the
individual ROIs and then use the analysis pipeline described in
Section~\ref{sec:sourcelist} to analyze those simulated maps.  For these
simulations we used the same time and energy ranges and the same ROI as
the analysis on the real sky, and we employed the Off.~IEM.

For the simulations described in this section we simulate blazars
isotropically distributed in the $40^{\circ}\times40^{\circ}$ ROI,
with fluxes taken from the $dN/dS$ derived
by~\citet{2010ApJ...720..435A} using the 1FGL
catalog~\citep{2010ApJS..188..405A}.  We simulated individual blazars
using a PL SED with $\Gamma$ extracted from a Gaussian distribution
with average 2.40 and standard deviation 0.30 as found for the
3FGL~\citep{2015ApJS..218...23A}.  We simulated blazars with 
an energy flux integrated between $0.3-500$ GeV of $>9\times 10^{-8}$ MeV cm$^{-2}$ s$^{-1}$
in order to have a sizable number of simulated sources below the detection threshold.

To model the Galactic disk and bulge PSR populations, ideally we would
start with a known luminosity function for PSRs, or derive one starting with the 210 publicly
announced $\gamma$-ray PSRs.   However, because of complications
including the incompleteness of the radio pulsar sample and variations in
detection efficiency across the sky, and since the PSR sample 
covers the entire sky well beyond our ROI, the PSR
luminosity function is poorly constrained and difficult to extract.
See, e.g.,~\citet{Strong:2006hf,Cholis:2014noa,Petrovic:2014xra,Bartels:2015aea} both for previous 
estimates of the luminosity function and for discussions of the complications.

Therefore we adopt a staged approach.  We first assume a PL
shape $dN/dL\propto L^{-\beta}$ for the luminosity function and
estimate the slope ($\beta$) from the data as described below.  We
then derive the normalization given that slope by using simulations
to estimate the number of $\gamma$-ray
pulsars that would be required to explain the GC excess and energy
flux distribution of 3FGL sources with curved SEDs using
simulations.  Finally, we reuse those simulations to derive the
efficiency for PSRs to pass our selection criteria.

Given for each PSR the energy flux $S$ and distance $d$ we calculate
the luminosity: $L = 4\pi S d^{2}$, where $S$ is 
 integrated in the energy range $0.3-500$ GeV,~\footnote[9]{Note that this
    differs from previous publications, which use 100~MeV as the lower
    bound of the integration range for the luminosity.} as derived with the analysis
  described in Section~\ref{sec:psrblazar} and the distance ($d$) is
  taken from the ATNF catalog version 1.54~\citep{2005AJ....129.1993M}  using the
continuously updated web page~\footnote[10]{We always use the
    {\it Dist\_1} parameter, namely the best distance estimate
    available, when it exists (see the ATNF catalog for more
    information:
    \url{http://www.atnf.csiro.au/people/pulsar/psrcat/}).}.  The
catalog provides distance measures for 135 out of 210 PSRs and from
these we can derive the observed luminosity distribution $dN/dL$. The
missing 75 pulsars are mostly {\it Fermi} $\gamma$-loud and
radio-quiet.  We also calculate the $dN/dL$ separately for young
PSRs and MSPs without correcting for the detection
efficiency.

\begin{figure}
	\centering
\includegraphics[width=1.03\columnwidth]{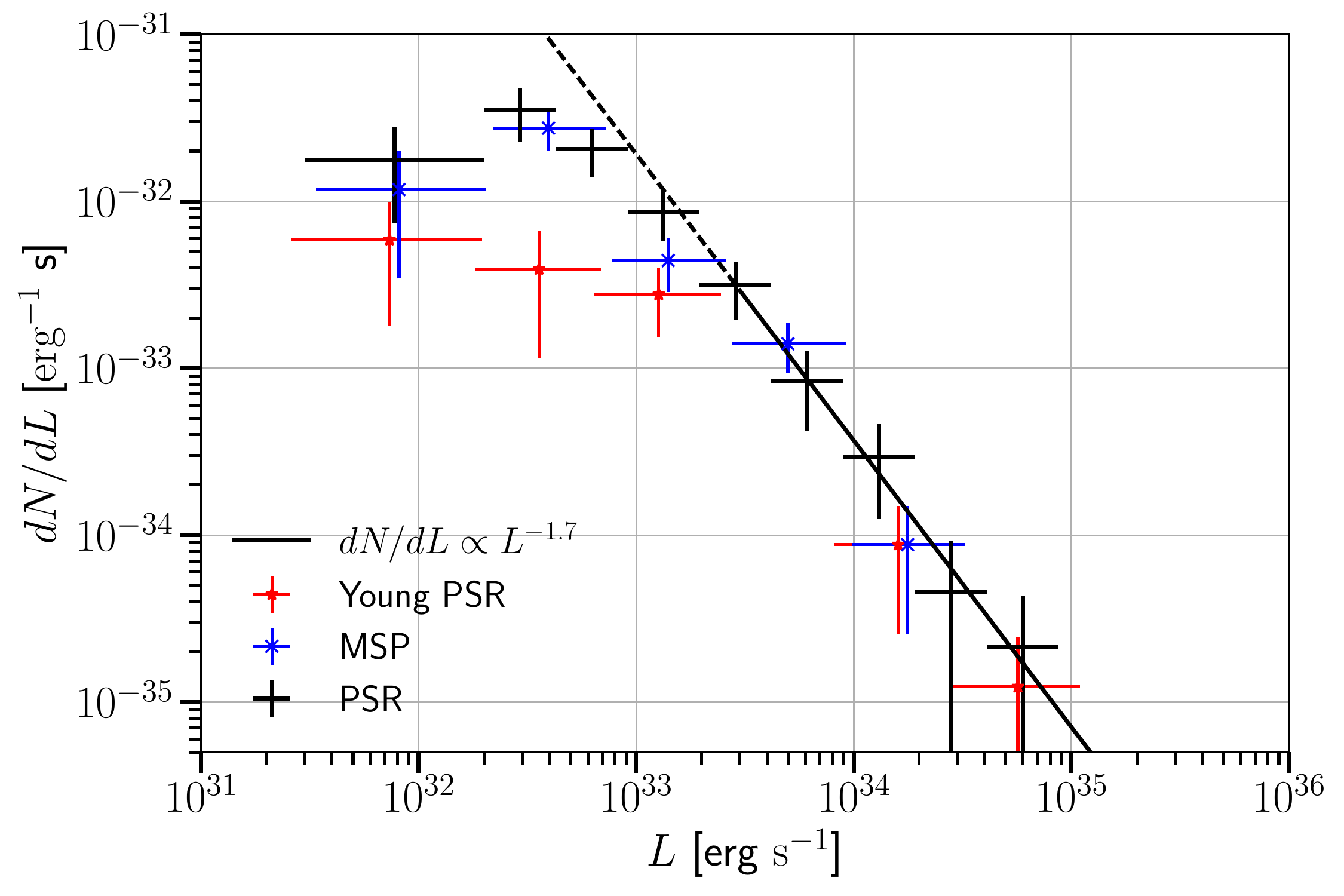}
\caption{Observed luminosities for young PSRs (red data),
  MSPs (blue data) and the whole population of PSRs with $d<1.5$ kpc (black data). The best fit
  to the luminosity distribution for $L> 3\times 10^{33}$ erg s$^{-1}$ is also reported (black line).  
  The luminosity is
  integrated over the energy range $[0.3,500]$ GeV.}
\label{fig:dNdLN} 
\end{figure}

Since the PSR sample detected by the LAT is known to be incomplete and
we do not correct for the detection efficiency, we select sources
within a distance of 1.5~kpc from the Earth.  Indeed, considering
luminosities in the range $[3\times 10^{33},10^{36}]$ erg s$^{-1}$ and
a distance of 1.5~kpc implies energy fluxes in the range
$[7 \times10^{-6},3\times 10^{-3}]$ MeV cm$^{-2}$ s$^{-1}$, for which
the LAT efficiently detects sources.  
In Figure~\ref{fig:dNdLN} we
show the luminosity distribution for our sample of PSRs with $d<1.5$
kpc.  We then perform a fit to the data starting from
$L=3\times 10^{33}$ erg s$^{-1}$ to avoid the change in slope at low
luminosities due to the incompleteness of the LAT detections at the
low-luminosity end. We use a PL shape $dN/dL\propto L^{-\beta}$ and
the fit yields $\beta=1.7\pm0.3$. 
Our fit differs from the data
points only below $10^{33}$ erg s$^{-1}$ where it is
difficult to identify PSRs with $\gamma$-ray data.  

  For reference, the faintest $\gamma$-ray pulsar is PSR J0437-4715 with
  $L=3.55\times10^{31}$ erg s$^{-1}$, and the faintest
  $\gamma$-ray pulsar found in a blind search is PSR J1741-2054 with
  $L=1.51\times10^{33}$ erg s${^-1}$.
Although the selection we applied in source luminosity and distance reduce significantly 
the incompleteness of pulsar sample,
the efficiency for the detection of PSRs for this distance and luminosity 
range might be smaller than 100\% and thus the intrinsic luminosity
distribution even steeper; however we will use the value of 
$\beta=1.7$ to examine the issue of how many PSRs would be required
to produce the GCE.

Our estimate of $\beta$ for PSRs with
$L_{\gamma}> 3\times 10^{33}$~erg s$^{-1}$ is similar to that found for
MSPs by~\citet{Cholis:2014noa, Hooper:2015jlu,Winter:2016wmy}.  In
these papers a break at around $10^{33}$ erg s$^{-1}$ or a slightly curved
luminosity function is considered.  However, since the slope of the
luminosity function is 1.7, the integrated luminosity is dominated by
the bright sources.  Therefore, a change of $dN/dL$ below $10^{33}$
erg s$^{-1}$ does not significantly affect our results.  We also point out
that~\citet{Winter:2016wmy} have estimated the completeness of the
Second {\it Fermi}-LAT catalog of pulsars~\citep{2013ApJS..208...17A}
finding that it is almost 100\% for pulsars with luminosity greater than
$10^{35}$ erg s$^{-1}$.  The least-luminous PSR detected is $3.55\cdot 10^{31}$
erg s$^{-1}$ while the most luminous is $10^{36}$ erg s$^{-1}$. We therefore
simulate luminosities between $10^{31}$ erg s$^{-1}$ and $10^{36}$ erg s$^{-1}$ to
include PSRs below the current LAT detection threshold.
Furthermore, throughout this paper, we quote the total
  number of PSRs with $L = [10^{33}, 10^{36}]$ erg s$^{-1}$ in the Galactic disk ($N_{\rm disk}$) 
  and bulge ($N_{\rm bulge}$) to specify the normalization of $dN/dL$.

\section{Simulating the Galactic PSR population}
\label{sec:gal_dist}

In this section we report our assumptions for the disk and Galactic bulge populations of PSRs
and explain how we simulate these two populations.

\subsection{Galactic Disk PSRs}
\label{sec:disk_dist}
\begin{figure}
	\centering
\includegraphics[width=1.03\columnwidth]{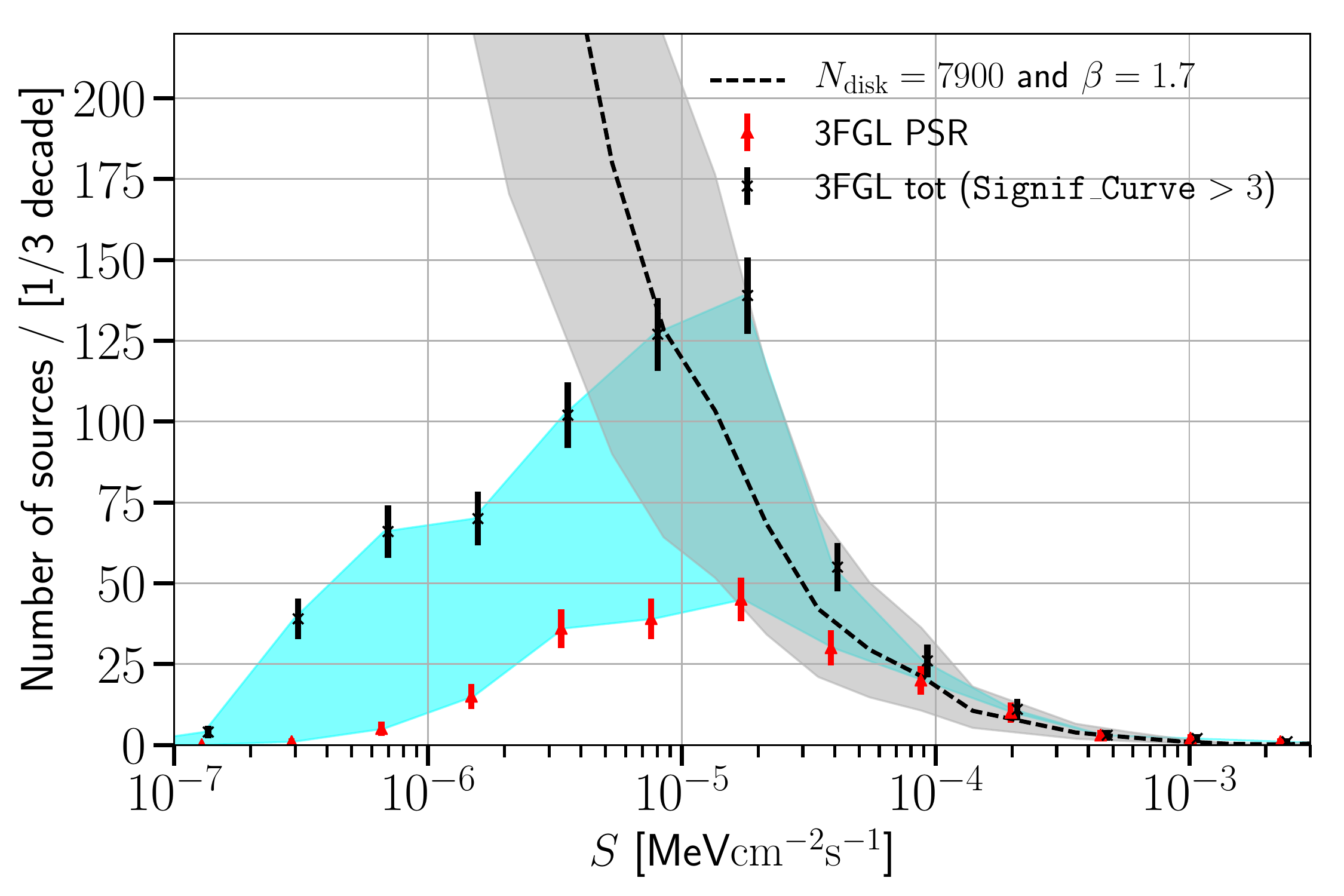}
\caption{Flux histogram of 3FGL PSRs alone (red triangles) or added to
  the flux distribution of unassociated 3FGL sources with curvature
  $\texttt{Signif\_Curve}>3$ (black points). The cyan band
  represents the region between the lower limit (already
  detected PSRs) and upper limit (3FGL PSRs plus unassociated 3FGL
  sources with detected spectral curvature). Finally the black curve
  (gray band) represents the benchmark (band between the minimum and
  maximum) number of disk PSRs.   The flux is integrated over the
  energy range $[0.3,500]$ GeV.}
\label{fig:Sdisk} 
\end{figure}

For our simulations we use the Galactocentric spatial distribution
$\rho(R)$ as modeled by~\citet{2004IAUS..218..105L}:
$\rho(R) \propto R^{n} \exp{(-R/\sigma)}$ with $n=2.35$ and
$\sigma=1.528$~kpc. The dependence on the distance from the Galactic disk is
modeled with an exponential cutoff $\rho(z)\propto \exp{(-|z|/z_0)}$ with
scale height $z_0=0.70$~kpc as in~\citet{Calore:2014oga}. The luminosity
function is modeled as a PL with index 1.70 over the range
$L=[10^{31},10^{36}]$ erg s$^{-1}$, see Section~\ref{sec:lumi_dist}.   

Analyses of the Galactic disk pulsar population estimate that it could
contain thousands of
objects~\citep{2013MNRAS.434.1387L,2013IAUS..291..237L,2004IAUS..218..105L}. These
estimates are derived from radio catalogs of pulsars, correcting
their spatial distribution for observational biases and using
information for the star formation rate and distribution in the
Galaxy.  

However, the radio and the $\gamma$-ray emission are only
  slightly correlated and many of the nearest radio pulsars are not
  detected by the LAT.  The current ATNF catalog lists 714 pulsars
  within 3~kpc of the Earth that have measured spin down energy loss
  rates ($\dot{E}$).  Of these, 257 have $\dot{E} > 10^{33}$ erg s$^{-1}$, the
  observed minimum for which pulsars emit $\gamma$
  rays~\citep{2016A&A...587A.109G}.  The LAT has detected about 30\% of
  these, most likely primarily due to differences 
  in radio and $\gamma$-ray emission beam solid angles of
  the source and to their distances.

In short, the overall number of $\gamma$-ray PSRs in the disk
population is not very well constrained.  A lower limit
is given by the identified $\gamma$-ray PSRs: in particular, for
fluxes integrated between $0.3-500$ GeV larger than $10^{-5}$
MeV cm$^{-2}$ s$^{-1}$ where the efficiency for the detection of PSRs is almost
$100\%$.  In Figure~\ref{fig:Sdisk} we
show the energy flux histogram for 3FGL PSRs.

This is, however, only a lower limit because many non-radio
PSRs may be present as sources in the 3FGL, but the pulsations have
not yet been detected in $\gamma$ rays. Without timing solutions
from radio observations, the detection of $\gamma$-ray pulsation is
challenging; see e.g.,~\citet{2011ApJ...742..126D} for a sensitivity
estimate.  To obtain an estimate of the upper limit of PSRs in the
disk we have selected the 3FGL unassociated sources with curvature
significance greater than 3.  We added their flux distribution to that
of the detected PSRs (Figure~\ref{fig:Sdisk}).  We expect that the
bright tail ($S>1.8 \cdot 10^{-5}$ MeV cm$^{-2}$ s$^{-1}$) of the flux
distribution for the disk population of PSRs should fall between the
already detected PSRs (111) and the sum of this with 3FGL unassociated
sources with $\sigma_{\rm{curv}}>3$ (237). This range is represented
by the cyan band in Figure~\ref{fig:Sdisk}.  
From this we estimate
that the Galactic disk PSR population consists of between 4000 and
16000 $\gamma$-ray emitting sources with $L  = [10^{33},10^{36}]$ erg s$^{-1}$ with $\beta=1.7$,
i.e., $N_{\rm disk} = [4000, 16000]$. This result is derived
with the Galactocentric spatial distribution as modeled
by~\citet{2004IAUS..218..105L} and it slightly depends on the
assumed radial distribution of PSRs in the disk.  The flux
distribution of this disk population is displayed with a gray band in
Figure~\ref{fig:Sdisk} and for energy fluxes larger than $10^{-5}$ MeV
cm$^{-2}$ s$^{-1}$ is perfectly consistent with the cyan band. 


\subsection{Galactic Bulge PSRs}
\label{sec:bulge_dist}

In this section, we consider the properties that would be needed for a 
Galactic bulge population of pulsars to generate the GC excess, and find
good agreement with previous works~\citep{2012PhRvD..86h3511A,
  2015ApJ...812...15B, Bartels:2015aea, Hooper:2015jlu}.

\begin{figure*}
	\centering
\includegraphics[width=1.03\columnwidth]{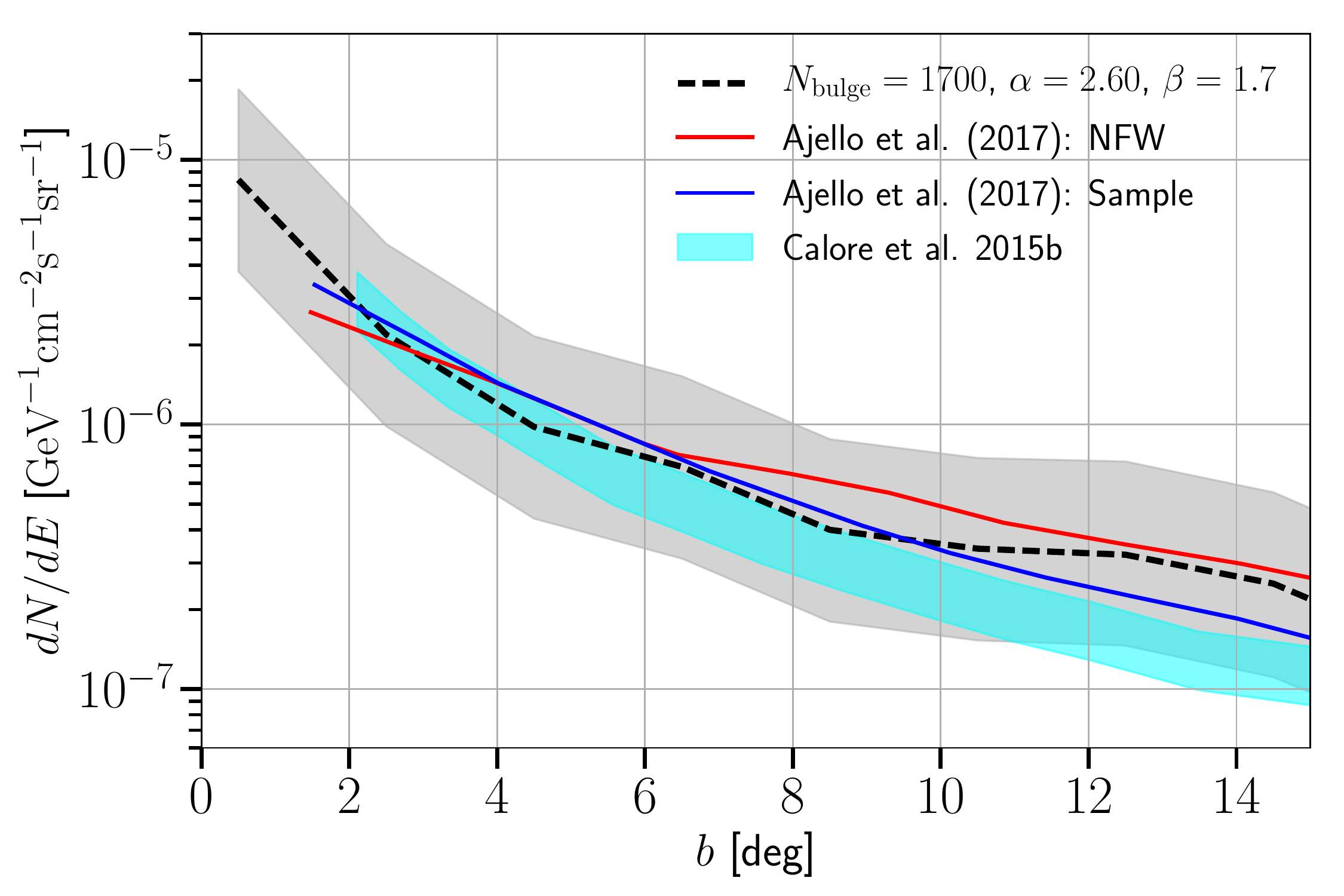}
\includegraphics[width=1.03\columnwidth]{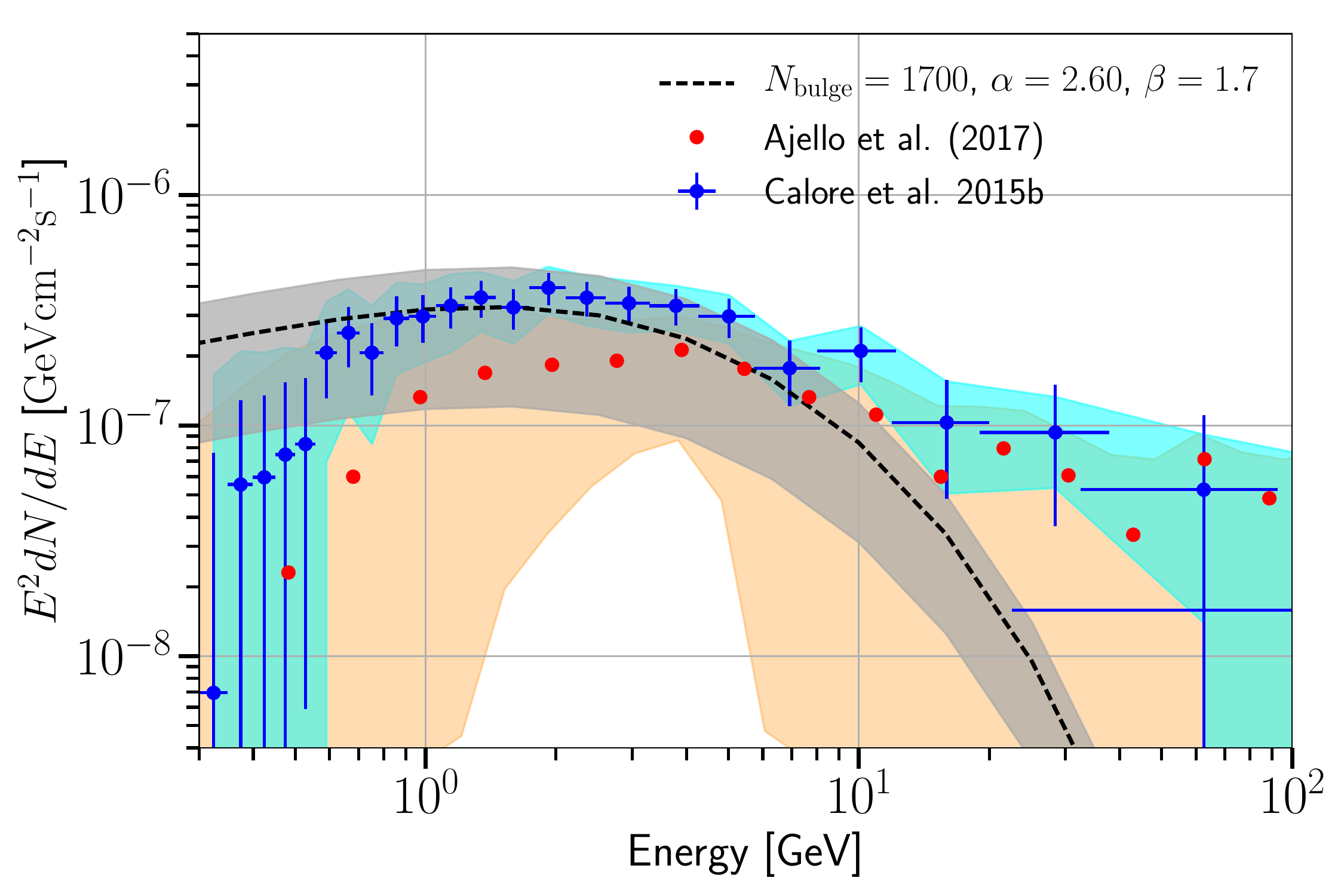}
\caption{Left panel: latitude profile of the intensity of the GC excess at 2~GeV
  and longitude $0^{\circ}$ for a simulation of a Galactic
    bulge PSR population that could reproduce the GC excess, i.e.,
    $N_{\rm{bulge}}=[800, 3600]$ for $L = [10^{33},10^{36]}$ erg s$^{-1}$} (gray band and black dashed line), and for the
  GC excess as found by~\citet{Calore:2014xka} (cyan band)
  and~\cite{TheFermi-LAT:2017vmf} (blue and red solid curves). Right panel:
  energy spectrum from a simulation of a Galactic
    bulge PSR population that could reproduce the GC excess, i.e., 
    $N_{\rm{bulge}}=[800, 3600]$ for $L = [10^{33},10^{36]}$ erg s$^{-1}$ (gray band and black dashed line), and for the
  GC excess as found by~\citet{Calore:2014xka} (cyan band and blue
  data) and~\citet{TheFermi-LAT:2017vmf} (orange band and red data).
\label{fig:bulgepsr} 
\end{figure*}

We model the spatial distribution of the Galactic bulge PSR
population as spherically symmetric with respect to the GC with a
radial profile $dN/dr \propto r^{-\alpha}$ for $r<3$ kpc and 0
elsewhere and with $\alpha=2.60$ in order to approximately match a
generalized NFW with slope of 1.3.  This spatial distribution is
consistent with the morphology of the GC
excess~\citep{Calore:2014xka,2016PDU....12....1D,2016ApJ...819...44A,
  TheFermi-LAT:2017vmf} and gives a latitude profile of the
$\gamma$-ray intensity from PSRs with a similar shape to the excess
(see left-hand panel Figure~\ref{fig:bulgepsr}).  As with the disk
population, we model the luminosity function as a PL with $\beta=1.7$
over the range $L=[10^{31},10^{36}]$ erg s$^{-1}$.  For each simulated
PSR we draw a location and luminosity from the relevant spatial
distribution and $\gamma$-ray PSR luminosity function and sample
values from the distributions of $\Gamma$ and
$\log_{10}(E_{\rm{cut}})$ given in the fifth row of
Table~\ref{tab:psrcandspectra}.  We then derive the SED of each PSR,
and simulate PSRs until their total energy spectrum is of the same
intensity as the GC excess as reported by~\citet{Calore:2014xka} and
\citet{TheFermi-LAT:2017vmf}.

In the left panel of Figure~\ref{fig:bulgepsr} we compare the average
latitude profile from 20 simulations with the intensity of the GC
excess and in the right panel we compare the total SED from simulated
PSRs and the GC excess spectrum as derived by~\citet{Calore:2014xka}
and~\citet{TheFermi-LAT:2017vmf}. The gray band in both plots is
derived from the possible range of source counts that this component
could contain and includes the reported systematic uncertainties of the latitude profile and energy spectrum
of the GC excess. This procedure finds that a Galactic bulge PSR
population capable of generating the GC excess would need to include
800--3600 sources with $L = [10^{33},10^{36]}$
erg s$^{-1}$, i.e., $N_{\rm bulge} = [800,3600]$.

\section{Conclusions}

\label{sec:conclusions}
We have analyzed 7.5 years of Pass 8 {\it Fermi}-LAT data for the
energy range $[0.3,500]$ GeV in a $40^{\circ}\times40^{\circ}$ around the GC in order to
provide a list of PSR candidates and test the pulsar interpretation of
the GC excess.  Employing two IEMs we detect about 400 sources, a
factor of about two more than in the 3FGL
catalog~\citep{2015ApJS..218...23A} and five more than in the
1FIG~\citep{2016ApJ...819...44A} (derived for $[1,100]$ GeV and for
$15^{\circ}\times15^{\circ}$); these latter analyses were based on
shorter time intervals of data than we consider here.  We then studied
the SEDs of $\gamma$-ray PSRs and 3FGL blazars using
a PLE shape and found that the distributions of photon index and
energy cutoff parameters for these two populations are very well
separated, with typical values of $\Gamma<2$ and
$E_{\rm{cut}}<10$~GeV for PSRs.  Moreover, about 82\%
of PSRs and only 9\% of blazars have $TS^{\rm{PLE}}_{\rm{curv}}>9$.
We thus use the selection criteria $TS^{\rm{PLE}}_{\rm{curv}}>9$,
$\Gamma<2$ and $E_{\rm{cut}}<10$~GeV to extract PSR
candidates from our seed list, finding 66 sources detected with both
IEMs.

We took the distribution of spectral parameters from the 210
identified $\gamma$-ray PSRs and the luminosity distribution of PSRs
within 1.5~kpc~\footnotemark[8].  We used parameters given by
~\citet{2004IAUS..218..105L, Calore:2014oga} to model the spatial
distribution of the disk population of PSRs.  With this model, we find
that given the number and distributions of unassociated 3FGL sources
with curved SED we constrain the number of Galactic disk PSRs to be in
the range $[4000,16000]$ for a luminosity function with slope 1.7 and
$L_{\gamma}=[10^{33},10^{36}]$ erg s$^{-1}$.  Similarly, we used the
latitude profile and energy spectrum of the GC
excess~\citep[e.g.,][]{Calore:2014xka,TheFermi-LAT:2017vmf} to model
the Galactic bulge PSR population and found that it must include 
800--3600 sources (most of them unresolved) if it is to explain the
GeV excess.

Facilities: \facility{Fermi}.


\begin{acknowledgements}
  The {\it Fermi} LAT Collaboration acknowledges generous ongoing
  support from a number of agencies and institutes that have supported
  both the development and the operation of the LAT as well as
  scientific data analysis. These include the National Aeronautics and
  Space Administration and the Department of Energy in the United
  States; the Commissariat \`a l'Energie Atomique and the Centre
  National de la Recherche Scientifique/Institut National de Physique
  Nucl\'eaire et de Physique des Particules in France; the Agenzia
  Spaziale Italiana and the Istituto Nazionale di Fisica Nucleare in
  Italy; the Ministry of Education, Culture, Sports, Science and
  Technology (MEXT), High Energy Accelerator Research Organization
  (KEK), and Japan Aerospace Exploration Agency (JAXA) in Japan; and
  the K. A. Wallenberg Foundation, the Swedish Research Council, and
  the Swedish National Space Board in Sweden.  Additional support for
  science analysis during the operations phase is gratefully
  acknowledged from the Istituto Nazionale di Astrofisica in Italy and
  the Centre National d'Etudes Spatiales in France.

  MDM and EC acknowledge support by the NASA Fermi Guest Investigator
  Program 2014 through the Fermi multi-year Large Program N. 81303
  (P.I. E.~Charles).

  We would like to thank Richard Bartels, Dan Hooper, Tim Linden,
  Siddhartha Mishra-Sharma, Nicholas Rodd, Benjamin Safdi and Tracy
  Slatyer for helping us to identify an error in the maximum
  likelihood analysis of the Galactic bulge and disk PSR populations 
  that was included in a previous version of this paper,
  (arXiv:1705.00009, v1).  Their work is described in a note will
  be posted on arXiv at the same time as this draft~\citep{bartels:XX}.

\end{acknowledgements}

\bibliography{paper}

\appendix

\section{A. Analysis pipeline and description of \fermipy tools}
\label{sec:fermipy}

We analyze each ROI with a pipeline based on the \fermipy
package and the {\it Fermi} Science Tools.  In the following description, we
denote with italics the \fermipy methods and configuration parameters
used in each step of the pipeline.

We start the analysis of each region with a model that includes 3FGL
sources with TS $> 49$, the IEM, and the isotropic template and begin
by optimizing the spatial and spectral parameters of this model.  We
first perform a global fit of the spectral parameters for all
components in the model.  For the global fit we retain the spectral
model (PL, LP or PLE) reported by the 3FGL.  We then relocalize all 3FGL
point sources using the {\it localize} method.  This method generates
a map of the model likelihood versus source position in the vicinity
of the nominal 3FGL position and finds the best-fit position and
errors by fitting a 2D parabola to the log-likelihood values in the
vicinity of the peak.  When localizing a source, we free the
normalization of the IEM and isotropic template and spectral
parameters of sources within 3$^\circ$ of the source of interest.
After relocalizing 3FGL sources, we repeat the global fit of the
spectral parameters of all components.

\begin{figure*}
	\centering
\includegraphics[width=0.495\columnwidth]{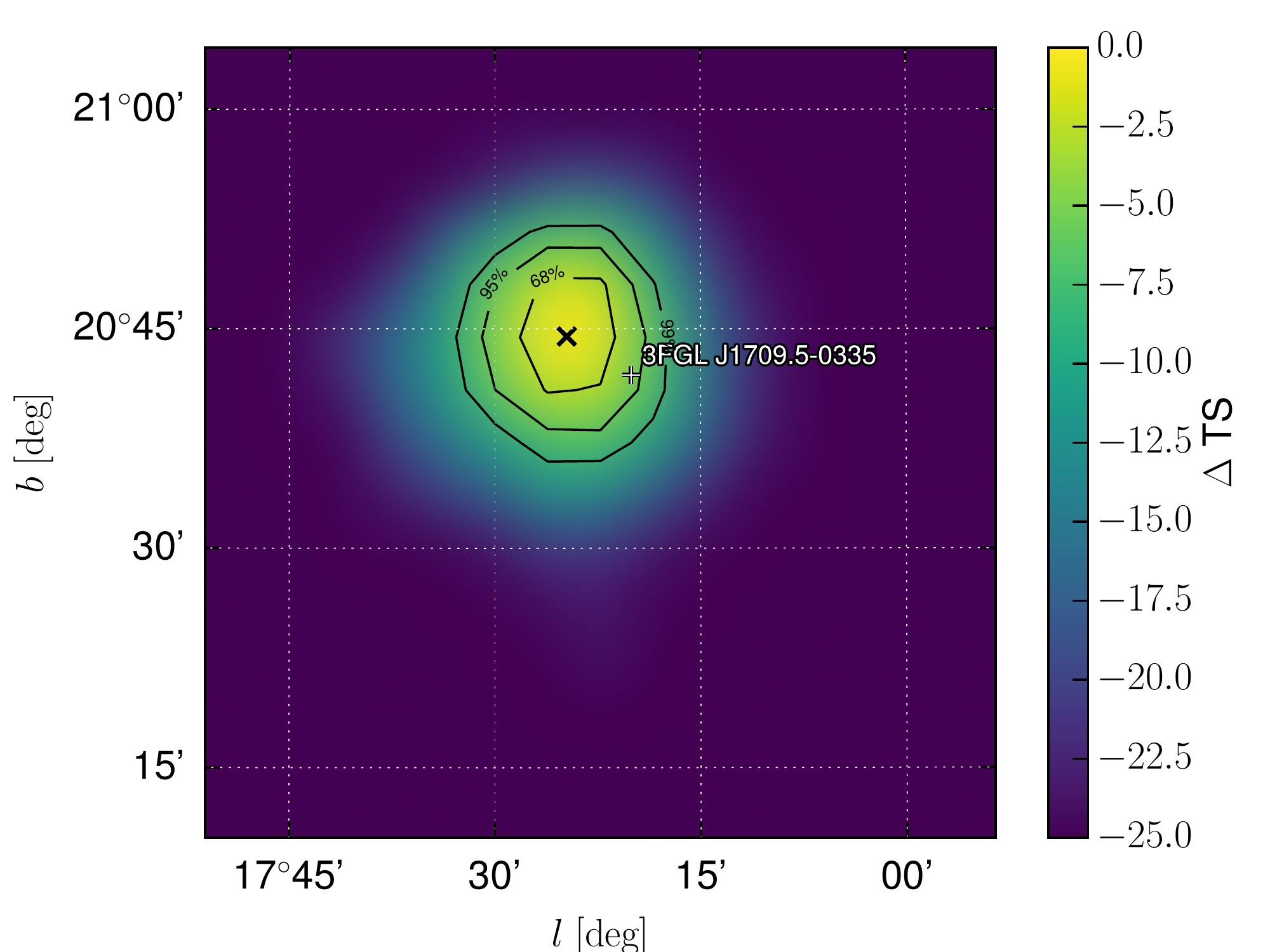}
\includegraphics[width=0.495\columnwidth]{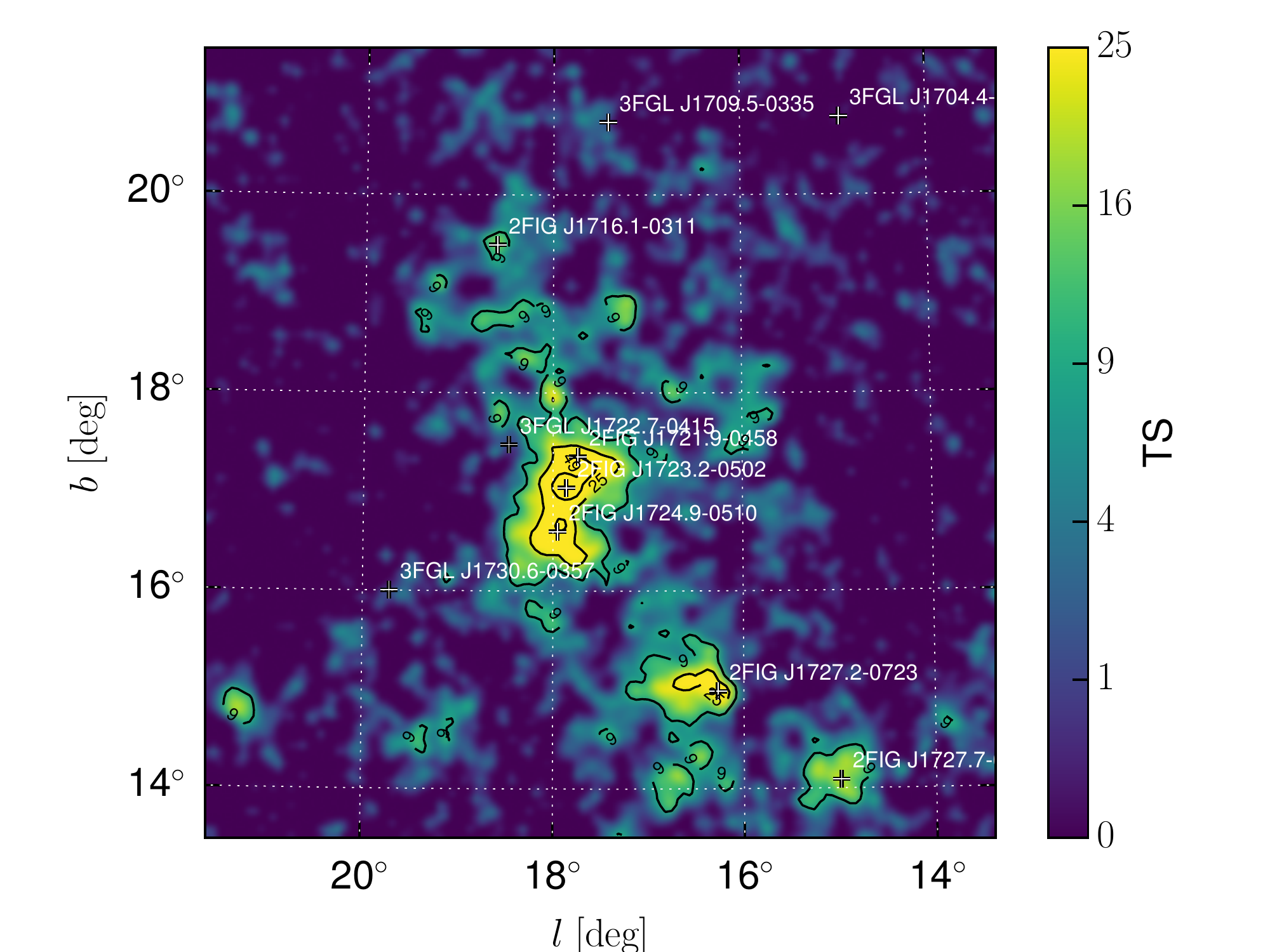}
\caption{Left panel: Positional uncertainty contours and best-fit
  position (solid black lines and x marker) from the localization of
  3FGL J1709.5$-$0335.  The color scale shows the difference in TS
  with respect to the best-fit position of the source.  The cross is
  the position of this source in 3FGL. Right panel: TS map of the
  ROI evaluated with a test source with a point-source morphology and
  a PL spectrum with photon index of 2.  The background model
  includes the IEM, isotropic template, and 3FGL sources with
  $TS>49$.}
\label{fig:reltsmap} 
\end{figure*}

On average, 3FGL sources move by $0.04^{\circ}$ in the relocalization
step. This is of the same order as the $68\%$ location uncertainty
radius for most 3FGL sources~\citep{2015ApJS..218...23A}. As an example, in
the left panel of Figure~\ref{fig:reltsmap} we show the result of the
relocalization for 3FGL~J1709.5$-$0335. The new position is offset by
$0.087^{\circ}$ with respect to the 3FGL position and the 3FGL $68\%$
positional uncertainty is $0.064^{\circ}$.

After relocalizing the 3FGL sources, we add new source candidates to
the model using the {\it find\_sources} method.  This method
iteratively refines the model by identifying peaks in a TS map of the
region with $\sqrt{TS} > $ {\it sqrt\_ts\_threshold} and adding a new
source at the position of each peak.  After each iteration a new TS
map is generated with an updated background model that incorporates
sources found in the previous iteration.  This procedure is repeated
until no peaks are found with amplitude larger than {\it
  sqrt\_ts\_threshold}.  To minimize the likelihood of finding
multiple peaks associated with a single source, the algorithm restricts
the separation between peaks found in an iteration to be greater than
{\it min\_separation} by excluding peaks that are within this distance
of a peak with higher TS.

We run {\it find\_sources} with a point-source test source model with
a PL spectrum and a fixed photon index of $\Gamma = 2$.  We use
{\it sqrt\_ts\_threshold} $=4$ and {\it
  min\_separation}$=0.4^{\circ}$.  The result is a list of source
candidates with $TS>16$.  On average we detected about six sources
with $TS>25$ per ROI.

In the right-hand panel of Figure~\ref{fig:reltsmap} we display an example TS map
derived prior to the source-finding step when only the IEM, isotropic
template and 3FGL sources with $TS>49$ are included.  From this figure
we note that a region with $TS>25$ is located in the center of
the ROI, near $(l,b)=(17.5^{\circ},17^{\circ})$ from which our
analysis extracted three new sources with $TS>25$.

\begin{figure}
	\centering
\includegraphics[width=0.55\columnwidth]{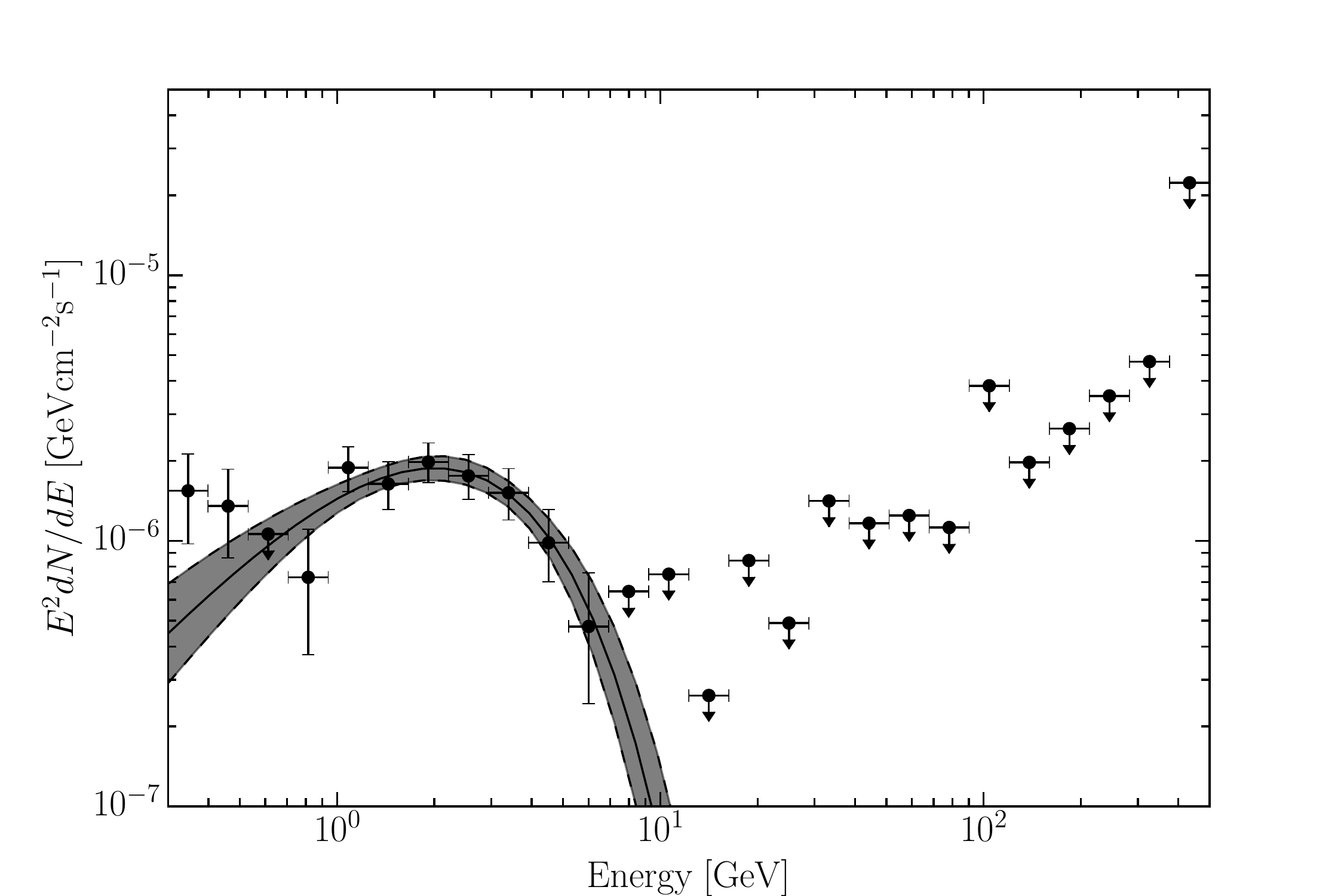}
\caption{SED of 3FGL J1730.6$-$0357 (black points) and best-fit PLE
  parameterization with $1\sigma$ uncertainty band (black line and
  gray band).}
\label{fig:sed} 
\end{figure}

We derive the SED for each source candidate in our list using the {\it
  sed} method. A likelihood analysis is performed in each
energy bin independently, using for the spectrum of the source a PL shape with a fixed
photon index of 2 and normalization free to vary. In this procedure we also
leave free the normalizations of the IEM, isotropic template, and the
normalizations of sources within $3^{\circ}$ of the source of
interest.  As an example, in Figure~\ref{fig:sed} the SED of 3FGL
J1730.6$-$0357 is reported together with a fit of a PLE
spectrum. 3FGL J1730.6$-$0357 has $TS^{\rm{PLE}}_{\rm{curv}}=40$
meaning that it has a significant curvature. The fit with a PLE in
fact gives a spectral index of 0.5 and energy cutoff of 1.4 GeV.

To avoid finding duplicate sources in regions where our ROIs overlap, 
we remove sources that are found in more than one ROI and 
that have an angular separation smaller than $0.2^{\circ}$.  Specifically 
we keep the version of the source that is closest to the center
of the ROI in which it was found.

\section{B. Minimum spanning tree source cluster-finding}
\label{sec:mst}

To avoid placing spurious point sources in regions where the IEM
under-predicted the Galactic diffuse emission we applied a source
clustering algorithm based on the minimum spanning tree (MST)
algorithm.  The MST algorithm calculates how to connect points with
minimum total length of the connections~\citep{10.2307/2033241}.  The
{\it fermipy.cluster\_sources} module applies the MST algorithm to
identify clusters of sources by joining sets of sources within a
maximum connection distance (the $dist$ parameter) and retaining those
clusters with at least a specified minimum number of sources (the
$nsrc$ parameter). We tested different values for $dist$ and $nsrc$.
We found that using $nsrc\ge4$ generally avoided creating spurious
clusters around chance spatial coincidences.  For $nsrc=4$, we
found that using $dist=0.6^{\circ}$ selected a few clusters that were
spatially associated with known regions of complicated Galactic
diffuse emission, such as supernova remnants and the GC.  Using larger
values, such as $dist=1.0^{\circ}$, resulted in clusters consisting of
chains of sources up to $\sim 4^{\circ}$ long in the Galactic plane.
Using $dist\le0.3^{\circ}$ and $nsrc=4$, resulted in no clusters being
found with either IEM.  The same four clusters found using the Off. IEM
and the values $dist=0.6^{\circ}$ and $nsrc=4$ were also found using 
$dist=0.7^{\circ}$, and only the cluster near the GC was found using    
$dist=0.5^{\circ}$.  In light of these studies, we adopted the value
$dist=0.6^{\circ}$ and $nsrc=4$ and obtain the results presented in 
Section~\ref{sec:sourcelist}.

\section{C. Source list and contents of FITS file}
\label{sec:app2fig}

Together with this paper, we are releasing the list of sources
detected in our analysis as a {\tt FITS} file.  The file contains a
single binary table with the source data.  The list includes 469
sources detected with $TS>25$ in a region with $|b|<20^{\circ}$ and
$|l|<20^{\circ}$.  The table has one row per source; the column names
and contents are described in Table~\ref{tab:sourcelist}.  When
applicable the units of the columns are given by the header keywords
following the {\tt FITS} standard. All of the spectral
  parameters are taken from the ROI optimization procedure described in
  Appendix~\ref{sec:fermipy}.

\begin{table*}
\center
\begin{tabular}{llcc}
Contents & Column Name & Units & Uncertainty \\
\hline
Source designation & {\tt Source\_Name} & \nodata & \nodata \\
Right ascension & {\tt RAJ2000} & [deg] & \nodata \\
Declination & {\tt DEJ2000} & [deg] & \nodata \\
Galactic longitude & {\tt GLON} & [deg] & \nodata \\ 
Galactic latitude & {\tt GLAT} & [deg] & \nodata \\
Containment radius ($68\%$) & {\tt pos\_68} & [deg] & \nodata \\
Containment radius ($95\%$) & {\tt pos\_95} & [deg] & \nodata \\
TS & TS & \nodata & \nodata \\
$TS^{\rm{PLE}}_{\rm{curv}}$ & {\tt TS\_curv} & \nodata & \nodata \\
Integrated photon flux between $E=[0.3,500]$~GeV & {\tt Flux300} & [ph cm$^{-2}$ s$^{-1}$] &  {\tt Unc\_F300} \\ 
Integrated energy flux between $E=[0.3,500]$~GeV & {\tt Energy\_Flux300} & [MeV cm$^{-2}$ s$^{-1}$] & {\tt Unc\_Energy\_Flux300} \\
Functional form of the SED & {\tt SpectrumType} & \nodata & \nodata \\
Spectral index & {\tt Spectral\_Index} & \nodata & {\tt Unc\_Spectral\_Index} \\ 
Cutoff energy (for PLE) & {\tt Cutoff} & [MeV] & {\tt Unc\_Cutoff} \\
Curvature parameter, $\beta$ (for LP) & {\tt beta} & \nodata & {\tt Unc\_beta} \\
IEM with which the source is detected & {\tt IEM} & \nodata & \nodata \\
Associated 3FGL source & {\tt 3FGL\_Name} & \nodata & \nodata \\
Classification of 3FGL source & {\tt 3FGL\_Class} & \nodata & \nodata \\
Cluster membership (Off. IEM) & {\tt Cluster\_Off} & \nodata & \nodata \\
Cluster membership (Alt. IEM) & {\tt Cluster\_Alt} & \nodata & \nodata \\
\end{tabular}
\caption{Contents of the 2FIG source list {\tt FITS} table.}
\tablecomments{When a source is detected with both of the IEM models the reported position, SED
  parameters as well as the photon and energy fluxes are the ones found
  with the Off. IEM.  We report the TS for curvature, and the SED
  parameters for the PLE only for PSR-like
  sources as defined in the main text.  
  For sources with a 3FGL association that was modeled with a
  LP spectrum we also report the curvature parameter, $\beta$.
  The {\tt IEM} column has value
  ``Off'', ``Alt'' or ``Off/Alt''.   The {\tt Cluster\_Off} and {\tt Cluster\_Alt} 
  columns give the index of the cluster to which a given source is associated, if any.}
\label{tab:sourcelist}
\end{table*}

\section{D. Generating simulated data with \fermipy}
\label{sec:appsim}
We use the {\it simulate\_roi} method to simulate the binned
$\gamma$-ray counts data in each ROI using the maximum-likelihood model of the
ROI.  Specifically, the method generates ``model cubes'' of the
expected number of $\gamma$-ray counts in each pixel and energy bin in
the ROI for the time interval of our analysis.  The method then
generates Poisson-distributed random numbers with expectation values
drawn from the model cube for each pixel and energy bin and produces a
simulated binned counts maps for each ROI.  This procedure results in
simulated $\gamma$-ray counts maps that are statistically identical to
those produced with {\it gtobssim}, which simulates individual
$\gamma$ rays and convolves them with the instrument response model.
The {\it simulate\_roi} method is many times faster than {\it
  gtobssim}, making the extensive simulations we have performed much
more tractable.

\section{E. Testing PSR selection criteria with simulated data}
\label{sec:testcurv}
As discussed in Section~\ref{sec:psrblazar}, $\sim$90\% of blazars in
the 3FGL catalog have an SED modeled with PL shape while the remaining
10\% are modeled with a LP.  On the other hand, about 82\% of PSRs
have energy spectra consistent with a PLE.  Employing the parameter
$TS^{\rm{PLE}}_{\rm{curv}}$ and making spectral fits of blazars and
PSRs with a PLE model we have shown that the criteria
$TS^{\rm{PLE}}_{\rm{curv}}>9$, $\Gamma<2.0$ and
$E_{\rm{cut}}<10$~GeV work very well to separate the PSR and
blazar populations.

\begin{figure*}
	\centering
\includegraphics[width=0.55\columnwidth]{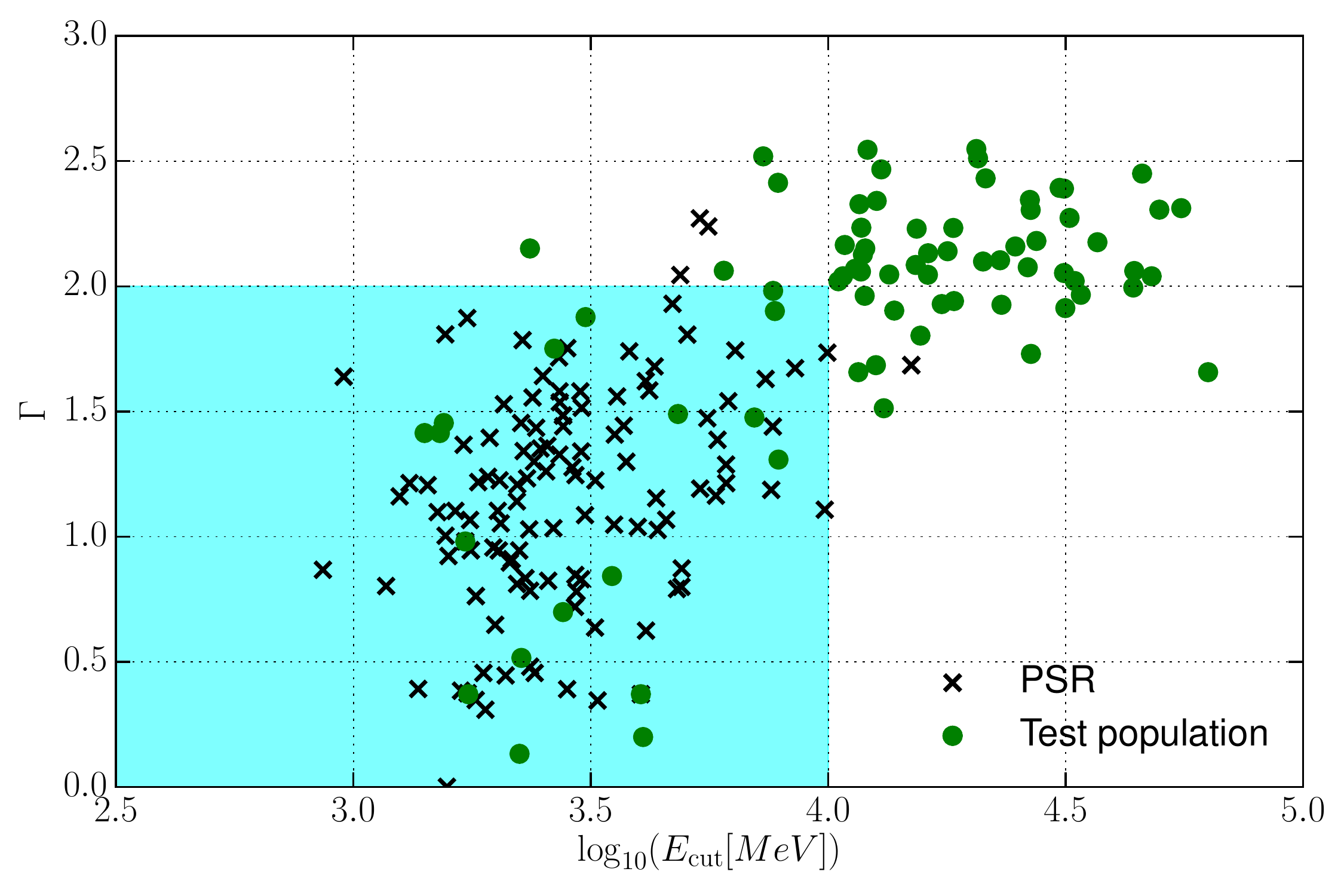}
\caption{Photon index $\Gamma$ and energy cutoff
  $E_{\rm{cut}}[\rm{MeV}]$ of PSRs (black points) and for a simulated 
  test population of sources isotropically distributed in the sky (red
  points). See the text for further details on the SED of these
  sources.}
\label{fig:psrcurved} 
\end{figure*}

In this Appendix we investigate how these criteria work for a
simulated test population of sources with curved spectra
compatible with a PLE, but with a slightly larger energy cutoff and
softer photon index with respect to PSRs. We want to test if this
additional population would severely contaminate our PSR candidates.
For this we simulate a bulge population of PSRs as explained in
Section~\ref{sec:bulge_dist} and 1500 sources with a photon index of $2.3\pm0.2$
and energy cutoff of $\log_{10}{(E_{\rm{cut}}[\rm{MeV}])} = 4.48\pm 0.25$
uniformly distributed in the GC region.  We choose these distributions
for the energy spectrum parameters to demonstrate that a putative
population of sources with a curved SED and with a distribution of
$\Gamma$ and $E_{\rm{cut}}$ that is fairly well separated from the
PSR-like criteria is not going to contaminate significantly our
selection of PSR-like sources because of mis-estimation of the 
spectral parameters.

We simulate fluxes of sources for this population from the source
count distribution of blazars derived by~\citet{2010ApJ...720..435A}.
We use the same analysis as used for the derivation of the source list
in the real sky.  For the SED we consider a PLE shape and evaluate the
best-fit parameters for $\Gamma$ and $\log_{10}(E_{\rm{cut}})$.  Then
we select the sources detected with $TS>25$ and
$TS^{\rm{PLE}}_{\rm{curv}}>9$.  The result for the values of $\Gamma$
and $\log_{10}(E_{\rm{cut}})$ is displayed in
Figure~\ref{fig:psrcurved}.  Also the detected sources satisfying
$TS^{\rm{PLE}}_{\rm{curv}}>9$ maintain a very good separation in the
$\Gamma-\log_{10}(E_{\rm{cut}})$ plane. 
Only $6\%$ of the non-PSR
sources detected with $TS>25$ have measured $\Gamma<2.0$ 
and $E_{\rm{cut}}<10$~GeV and $TS^{\rm{PLE}}_{\rm{curv}}>9$.
This result means that the presence of a putative source population
with an SED modeled with a PLE but with a softer photon index and
higher-energy cutoff would produce a contamination that is small with
respect to our PSR candidates.

\end{document}